\documentclass[twocolumn,english,aps,prb,showpacs,byrevtex,amsmath,amssymb,superscriptaddress]{revtex4-1}
\usepackage{fancyhdr}
\usepackage{amsmath}
\usepackage{amsfonts}
\usepackage{amssymb}
\usepackage{derivative}
\usepackage{subfigure}
\usepackage{epsfig}
\usepackage{pifont}
\usepackage{textcomp}
\usepackage{gensymb}
\usepackage{pifont}
\usepackage{enumerate}
\usepackage{color}
\usepackage{float}
\usepackage{bm}
\usepackage{ulem}
\begin{document}

\title{Mott-insulator Ca$_2$RuO$_4$ under a static external electric field}

\author{Giuseppe Cuono }

\affiliation{International Research Centre Magtop, Institute of Physics, Polish Academy of Sciences,
Aleja Lotnik\'ow 32/46, PL-02668 Warsaw, Poland}

\author{Carmine Autieri}
\email{autieri@magtop.ifpan.edu.pl}
\affiliation{International Research Centre Magtop, Institute of Physics, Polish Academy of Sciences,
Aleja Lotnik\'ow 32/46, PL-02668 Warsaw, Poland}

\date{\today}
\begin{abstract}
We have investigated the structural, electronic and magnetic properties of the Mott-insulator Ca$_2$RuO$_4$ under the application of a static external electric field in two regimes: bulk systems at small fields and thin films at large electric fields. Ca$_2$RuO$_4$ presents an S- and L-Pbca phase with short and long c lattice constants and with large and small band gaps, respectively.
Using density functional perturbation theory, we have calculated the Born effective charges as response functions. 
Once we break the inversion symmetry by off-centering the Ru atoms, we calculate the piezoelectric properties of the system that suggest an elongation of the system under an electric field. 
Finally, we investigated a four unit cells slab at larger electric fields and we found insulator-metal transitions induced by the electric field. By looking at the local density of states, we have found that the gap gets closed on surface layers while the rest of the sample is insulating. Correlated to the electric-filed-driven gap closure, there is an increase in the lattice constant c. Regarding the magnetic properties, we have identified two phase transitions in the magnetic moments with one surface that gets completely demagnetized at the largest field investigated. In all cases, the static electric field increases the lattice constant c and reduces the band gap of Ca$_2$RuO$_4$ playing a role in the competition between the L-phase and the S-phase.

\end{abstract}

\maketitle

\section{Introduction}







The study of compounds under the application of an external electric field has recently aroused great interest. One of the most important phenomena induced by the electric field is the control of the electronic properties of the systems. Among the several cases, recently great attention was devoted to the manipulation of the insulator-metal Mott transition\cite{Imada98} via an external electric field. The control of the Mott transition can be useful for electronic devices as for example resistance RAM \cite{Meijer08, Waser07}. The application of the electric field is complementary to the application of pressure, with the important difference that pressure influences the electronic states by modifying the structural parameters, while the electric field directly controls the electronic states, with many technological advantages.
A large electric field can control the carrier density in a region of an insulator, this is named as electrostatic carrier doping \cite{Rozenberg04,Inoue08}.

Furthermore, an electric field can also break the inversion symmetry. As a consequence of the breaking of the inversion symmetry, in theoretical models the hamiltonian is more anisotropic\cite{vanthiel2020coupling} and other terms in the hamiltonian are allowed as the spin-orbit Rashba\cite{PhysRevMaterials.3.084416,Rashba22} and the orbital Rahsba\cite{vanThiel2021coupling}. 
We have to mention that the breaking of the inversion symmetry can be introduced in many different ways, not only with an external electric field, but also in presence of surface, interfaces or inclusions\cite{PhysRevB.89.075102,PhysRevB.85.075126}. In particular, the interface between ferroelectric and magnetic materials was widely investigated in the last decade\cite{Amitesh2014APL,Autieri2014NJP,Hausmann2017}. 

\begin{figure}[h]
	\centering
	\includegraphics[width=8.6cm,angle=0]{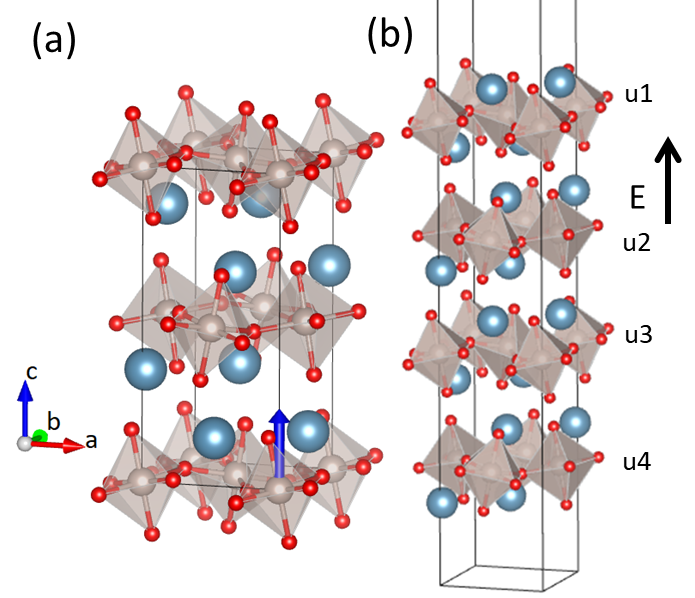}
	\caption{a) Crystal structure of the CRO bulk. Grey, red and blue spheres indicate the Ru, O and Ca atoms,
respectively. The blue arrow indicates the direction of the displacements of the Ru atoms when we break the inversion symmetry. b) The four unit cells slab. The black arrow indicates the direction of the applied electric field perpendicular to the surface. With u1, u2, u3 and u4 we indicate the four unit cells.}
	\label{crystal_structure}
\end{figure}

Ca$_2$RuO$_4$ (CRO) is a system that lends itself to analysis in the electrical field for its many phases and states \cite{Nakatsuji00,Cuoco06,Forte10,Pincini19} and where the scale associated with the Mott transition at T$_{MI}$=357 K is much greater than that associated with antiferromagnetism at T$_{N}$=110 K. 
The magnetic and electronic properties of CRO are sensitive to the coupling of spin, charge and the orbital degrees of freedom \cite{Koga04,Das18}.
It presents a Mott metal-insulator phase transition at T$_{MI}$=357 K from a low-temperature Mott-insulating phase to a high-temperature metallic-phase \cite{Alexander99}. This transition is accompanied by a structural transition, namely the compound has a small c-axis in the low-temperature phase, called S-Pbca, and a longer $c$ lattice constant in the high-temperature phase, therefore named L-Pbca configuration, both S- and L-Pbca configurations are orthorombic.
We know from the literature that the L-Pbca phase is metallic and the S-Pbca phase is insulating and the lower energy orbital xy is full \cite{Gorelov10,Zhang17}.
The unit cell contains four formula units with the RuO$_6$ octahedra settled in corner-shared planes alternated by CaO layers as shown in Fig. \ref{crystal_structure}.
It was shown that the Mott transition occurs because of the structural transition \cite{Gorelov10}, and the occupation of the orbitals changes from a configuration with xz and yz occupied to one with xy occupied, bringing the system to an orbitally ordered state at low temperature\cite{Okazaki13,Porter18}.\\

The Mott transition happens in the paramagnetic phase of the material at T$_{MI}$, however, it was found that insulator-metal transitions in CRO can be obtained under electric fields \cite{Nakamura13} or currents \cite{Zhang19,Cirillo19,Mattoni20} at lower temperature. 
Recently, it was also shown a way to induce a pattern formation by means of an applied electric field in CRO \cite{Gauquelin22}.
The electric field in this compound can modify the lattice constant c producing a competition between structural phases with different c lattice constant. In this paper, we want to focus on the effect on the electric field on the experimental insulating S-Pbca phase and on the hypothetical insulating L-Pbca phase as a benchmark. The latter cannot be observed experimentally without electric field, but the insulating L-Pbca could be stabilized by the electric field in future experiments.
\\

\subsection{Paper organization}	

In this paper, we study the structural, electronic and magnetic properties of Ca$_2$RuO$_4$ under the application of an external electric field with a theoretical computational analysis by using {\it ab-initio} density functional theory (DFT) method. We analyze both the regions of small and great electric fields applied. In the first region, we focus on the bulk and we calculate the response functions, while in the second one we build a four unit cells slab and we investigate the local density of states, the equilibrium c-axis, the band gap and the magnetic moments of the Ru atoms. The paper is organized as follows: in the next Section, we report the computational details, while in the third Section we present our results. Finally, in the last Section, we draw our conclusions.

\section{Computational details}

Our DFT simulations have been done by employing the Vienna ab initio simulation package (VASP) \cite{Kresse93,Kresse96a,Kresse96b}.
The projector augmented wave (PAW) \cite{Kresse99} technique has been used for the core and the valence electrons, with a cutoff of 480 eV for the plane-wave basis.
The calculations have been performed with an 11$\times$11$\times$4  k-points grid for the bulk and a 14$\times$14$\times$1  k-points grid for the slab, all centered in $\Gamma$.
The Local Density Approximation is enough to describe the metallic phases of ruthenates.
For the treatment of exchange-correlation, the Perdew-Burke-Ernzerhof (PBE) \cite{Perdew08}  generalized gradient approximation (GGA) has been used, and we have also considered the correlations for the Ru-4$d$ states by using a Coulomb repulsion U=3 eV on the Ru atoms in the antiferromagnetic insulating state \cite{Autieri_2016}.
For the Hund coupling, we have used a value in agreement with the literature for the $4d$/$5d$-electrons \cite{Vaugier12}, namely J$_H$ = 0.15 U has been employed. 
The experimental lattice constants are a$_{short}$=5.3945 {\AA}, b$_{short}$=5.5999 {\AA}, c$_{short}$=11.7653 {\AA} in the S-Pbca phase and a$_{long}$=5.3606 {\AA}, b$_{long}$=5.3507 {\AA}, c$_{long}$=12.2637 {\AA} in the L-Pbca phase \cite{Friedt01}.
For the calculation of the Born effective charges $Z^{*}$ and the piezoelectric tensor in the bulk, we have used the modern theory of polarization \cite{Resta92,Smith93} and the self-consistent response to finite electric fields\cite{Nunes01,Souza02} for bulk systems as implemented in VASP. For the thin film, we have constructed 4 unit cells along the (001) direction and we have added the electric field in the direction (001) perpendicular to thin film surfaces performing the same strategy used for several 2D materials in electric field\cite{Islam21,Islam22}. In the case of the slab, we have considered the dipole corrections to the potential as implemented in VASP in order to avoid interactions between the periodically repeated images \cite{Neugebauer92}.
Density functional theory does not fully reproduce the properties of the Mott-insulator at high temperature, therefore deviations from the DFT results are expected once we would include many-body and dynamical effects in the self-energy, which is a relevant property of many-body systems, especially in the non-magnetic phase and close to the Mott transition at T$_{MI}$.

\section{Results}

We divide our results into three subsections. In the first subsection, we report the investigation of the bulk with inversion symmetry. In the second one, we calculate the piezoelectric tensor after breaking the inversion symmetry by shifting the positions of the Ru atoms along the z-axis.  The direction of the displacements of the Ru is shown in Fig. \ref{crystal_structure}a). In the third subsection, we analyze the properties of a four unit cells slab, shown in Fig. \ref{crystal_structure}b), without and with the application of an external electric field.

\subsection{Properties of the bulk with inversion symmetry}

\begin{figure}[h]
	\centering
	\includegraphics[width=4.8cm,angle=270]{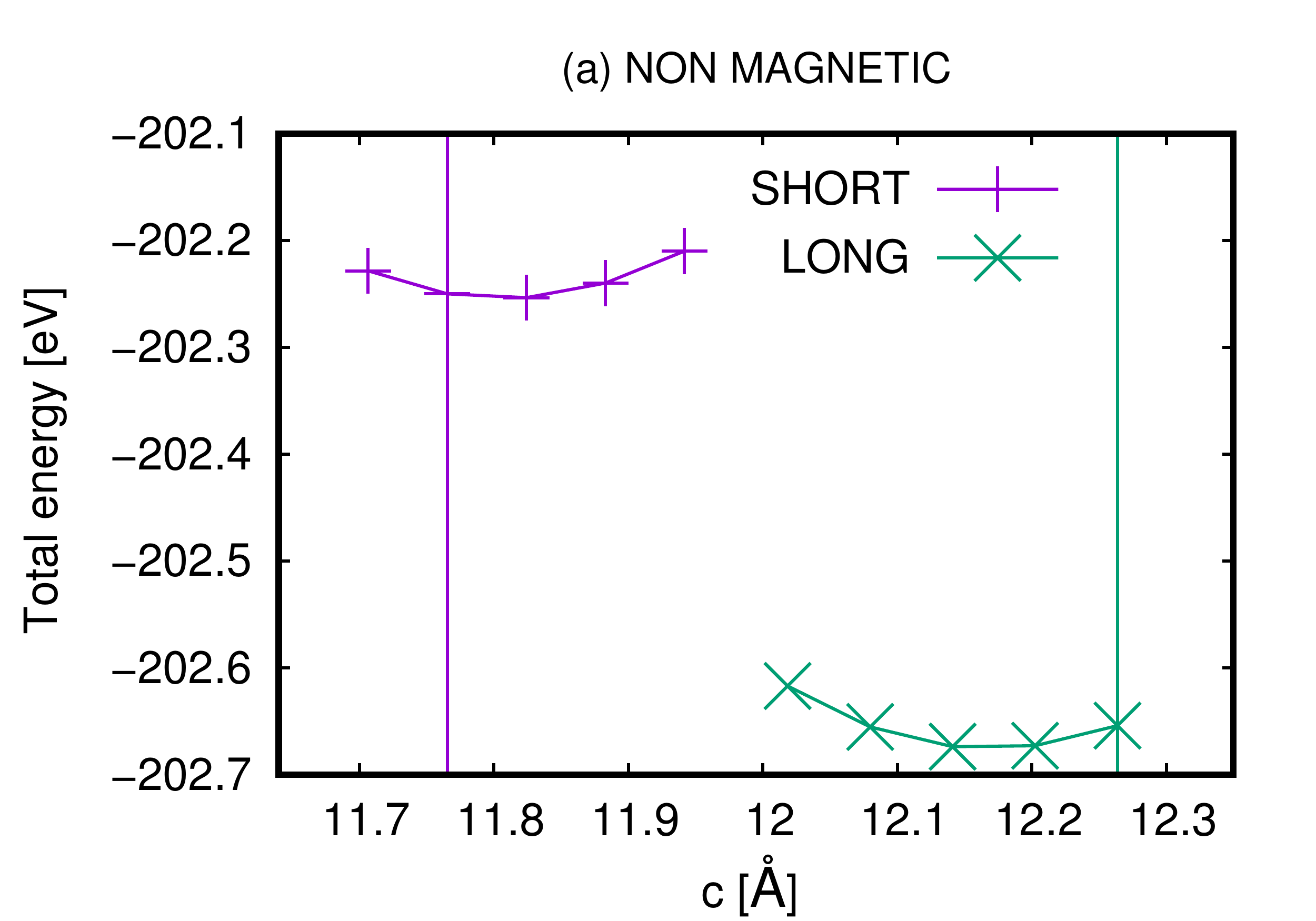}
\includegraphics[width=4.8cm,angle=270]{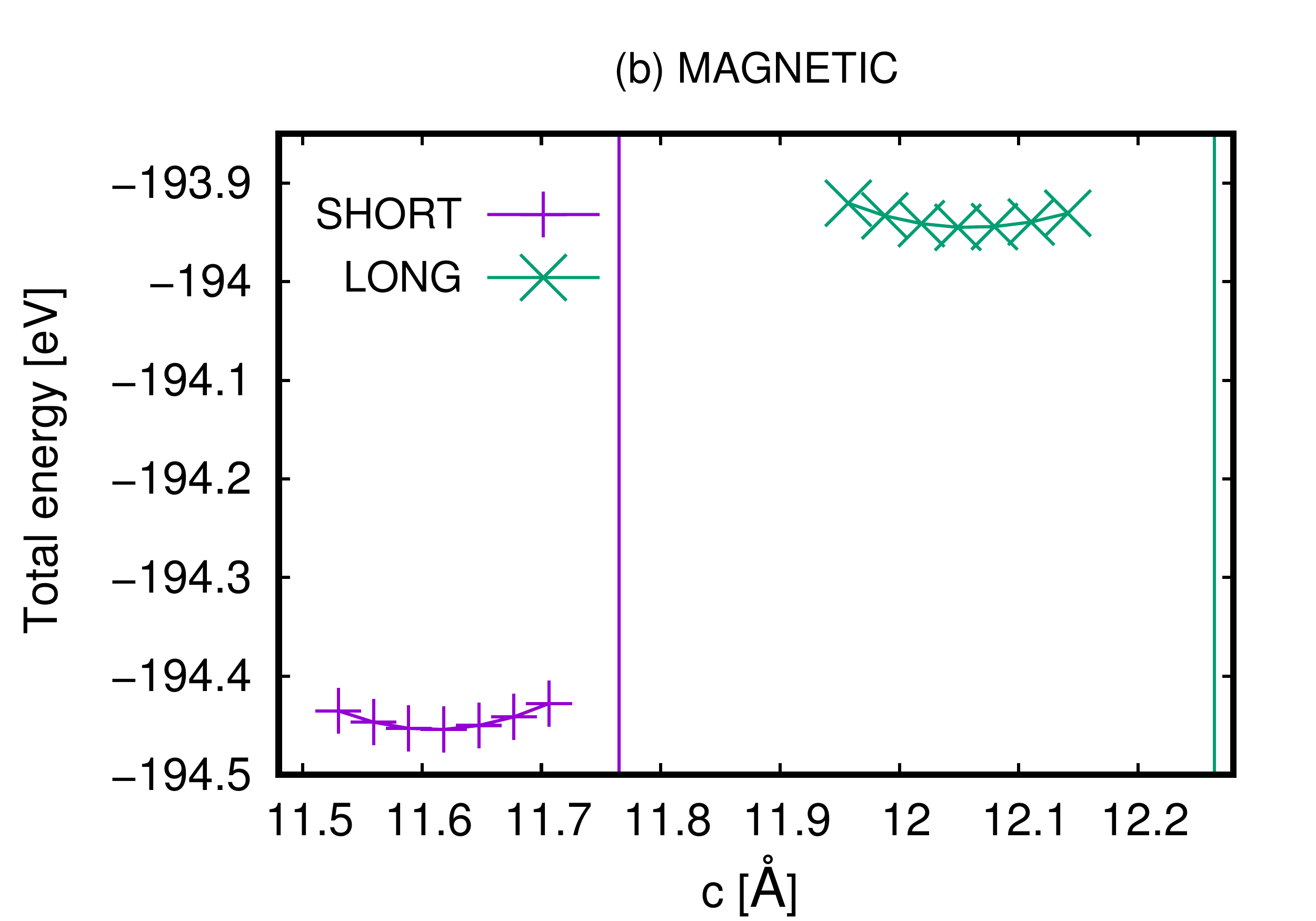}
	\caption{Total energy of the system as a function of the lattice constant c in the a) non-magnetic and b) in the magnetic case for the L- and S-Pbca phase. The in-plane lattice parameters have been fixed to the experimental values \cite{Friedt01} reported in the computational details.}
	\label{Ca2RuO4_Energy}
\end{figure}

In this subsection, we analyze the Ca$_2$RuO$_4$ bulk without and with the application of an external electric field.
First, we have compressed and elongated the system along the c-axis both in the S- and  L-Pbca phases and we have investigated how the energy of the compound varies as a function of the lattice constant c.  We have taken into consideration both the non-magnetic and the magnetic cases. 
In Fig. \ref{Ca2RuO4_Energy} we report the energy of the S- and  L-Pbca phases as a function of the c-axis, in panel a) in the nonmagnetic phase, while in panel b) in the magnetic phase. In the latter case, the Ru atoms of the primitive cell are in the checkerboard antiferromagnetic configuration. 
In both cases, the L-Pbca phase has a theoretical value of the c lattice constant larger than the S-phase. The difference between the theoretical and experimental c lattice constants is of the order of 1\% for the non-magnetic phases and f the order of 2\% for the magnetic phases. 
The non-magnetic phases are metallic, while the antiferromagnetic phases are insulating.
Even without considering dynamical effects, we obtained that in the metallic phase the L-Pbca phase is the ground state, while in the insulating phase the S-phase is the ground state in agreement with the experimental results. 
In the bulk, the band gaps of the antiferromagnetic phases are 0.88 eV  and 0.72 eV for the S- and L-Pbca phases, respectively.
\\

\noindent 
\begin{table*}[t!]
\begin{centering}
\begin{tabular}{|c|c|c|c|c|c|c|c|c|c|}
\hline 
Ion & $Z^{*}_{xx}$ & $Z^{*}_{xy}$ & $Z^{*}_{xz}$ & $Z^{*}_{yx}$ & $Z^{*}_{yy}$ & $Z^{*}_{yz}$ & $Z^{*}_{zx}$ & $Z^{*}_{zy}$ & $Z^{*}_{zz}$ \tabularnewline
\hline 
Ca  & 9.331  & -0.089  & 0.087  & -0.138  & 9.288  & -2.594  & 0.172  & 0.088  & 8.875\tabularnewline
\hline
O$_{1}$  & -4.794  & -1.101  & 0.128  & -1.003  & -4.656  & -1.471  & 0.129  & 0.054  & -3.558\tabularnewline
\hline
O$_{2}$ & -4.558  & -0.017  & 0.383  & -0.299  & -4.651  & -1.355  & 0.227 & 0.126 & -5.915\tabularnewline
\hline
Ru$_{\uparrow}$  & 0.045  & -0.464   & -0.119  & 0.412  & 0.037  & 3.121 & -0.409 & 0.377 & 1.193\tabularnewline
\hline
Ru$_{\downarrow}$ & 0.045  & 0.464   & 0.119  & -0.412  & 0.037  & 3.121 & 0.409 & 0.377 & 1.193\tabularnewline
\hline 
\end{tabular}
\par\end{centering}
\caption{$Z^{*}$ tensors for the bulk in the S-phase. O$_{1}$ and O$_{2}$ are the basal and the apical oxygens, respectively. Ru$_{\uparrow}$ and Ru$_{\downarrow}$ are the two Ru with opposite magnetic moments since we are considering the checkerboard antiferromagnetic configuration. }
\label{BEC} 
\end{table*}

\begin{figure*}[ht!]
\centering
\includegraphics[width=4.8cm, angle=270]{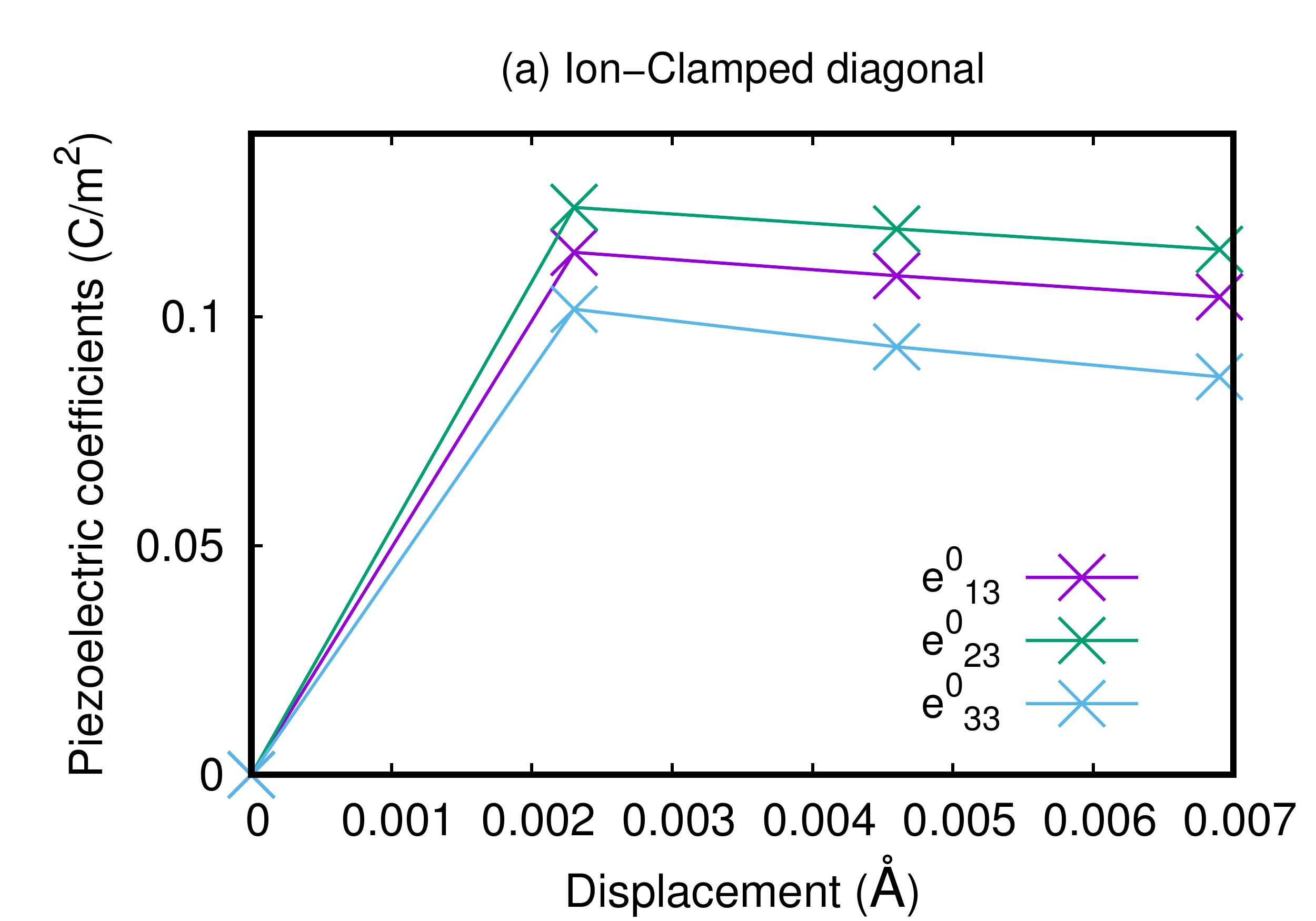}
\includegraphics[width=4.8cm, angle=270]{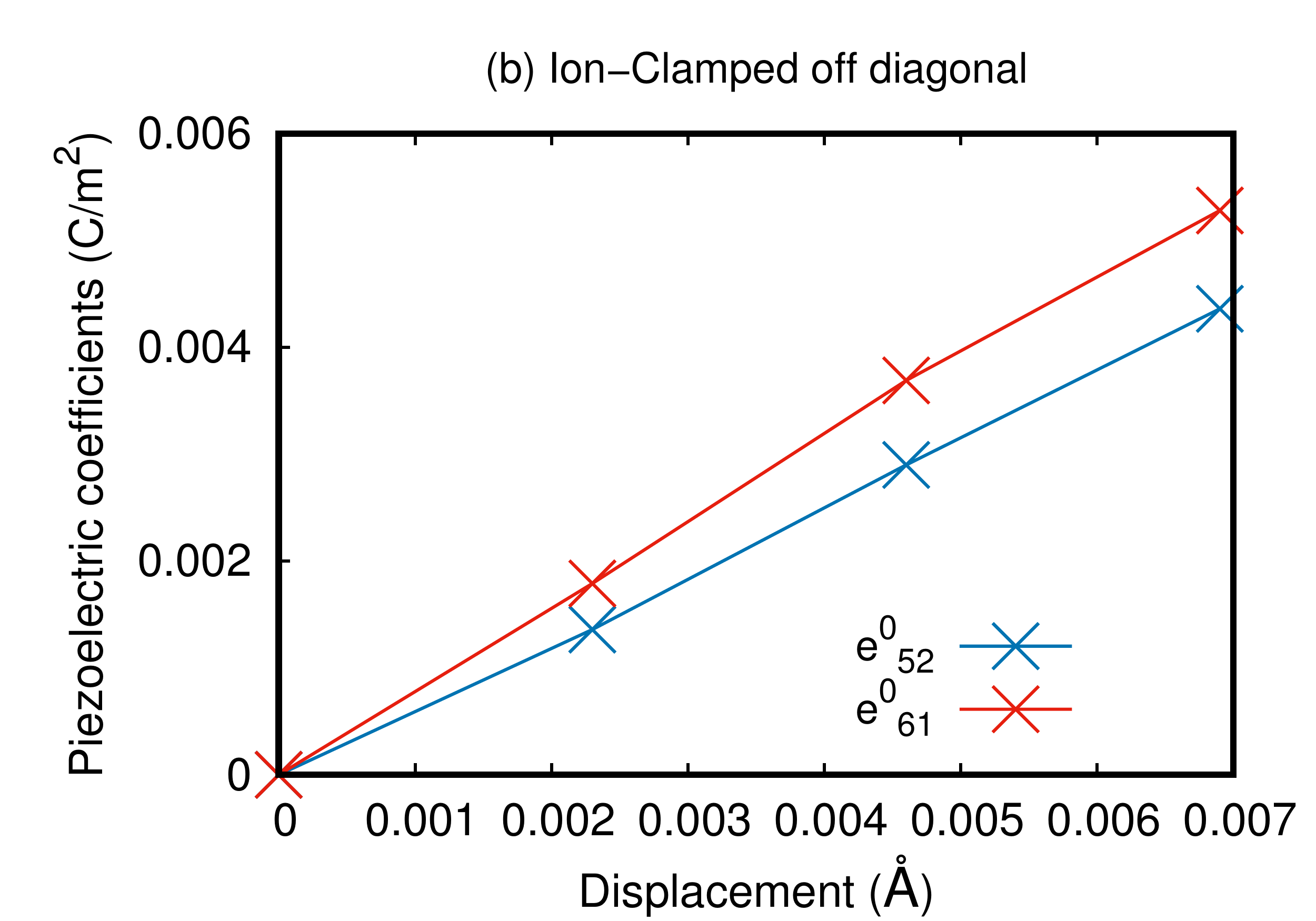}
\includegraphics[width=4.8cm, angle=270]{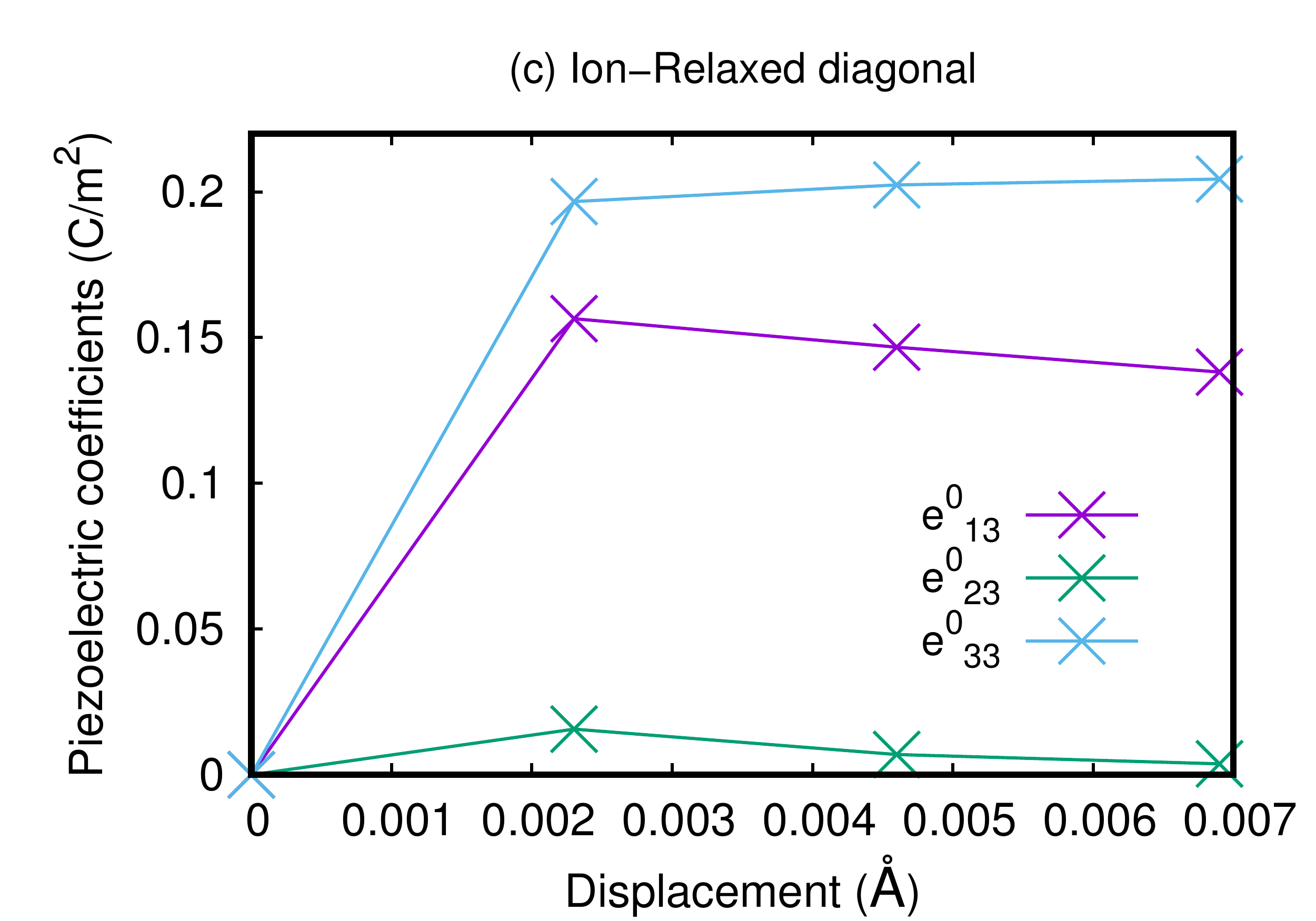}
\includegraphics[width=4.8cm, angle=270]{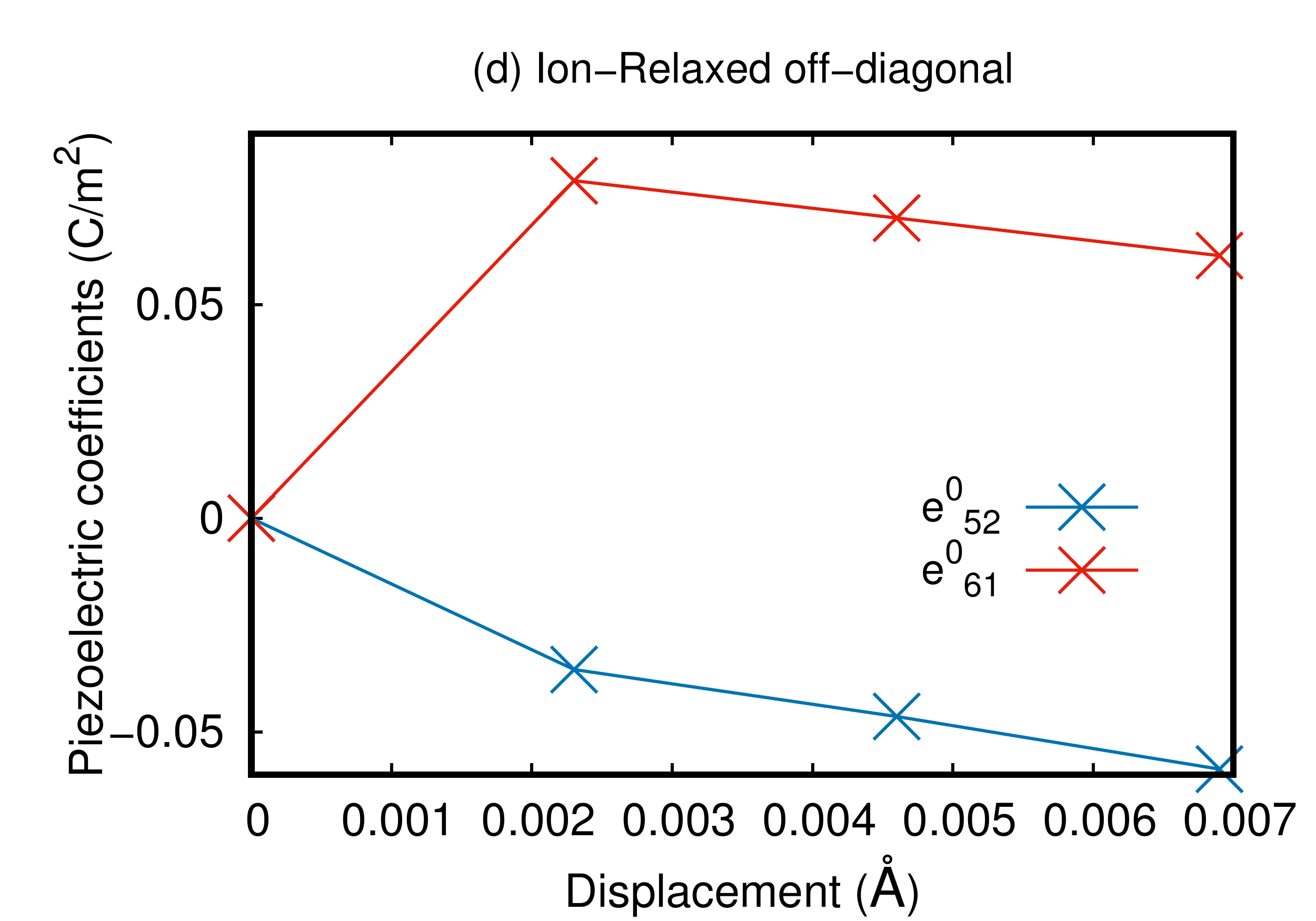}
\caption{Evolution of the piezoelectric components for the S-phase as a function of the displacement of the Ru atoms along z-axis for the bulk. We report the ion-clamped and ion-relaxed cases (diagonal and off-diagonal refer to the components of the stress tensor) in panels (a)-(d).}
\label{Piezoelectric tensor}
\end{figure*}

Then, we applied an electric field along the z direction for the S-Pbca phase, with qE=5$\times$10$^{-4}$ eV/{\AA}=5$\times$10$^4$ eV/{cm}, where E is the electric field and q is the elementary charge, and we have calculated the response functions of the system, like the Born effective charges $Z^{*}$. For larger electric fields, the numerical simulations do not converge because we are close to the onset of Zener tunneling\cite{Souza02}.
Under electric fields of this order of magnitude, the system is an insulator and the gap is 0.60 eV for the short crystal structure. Therefore, even a relatively small electric field tends to close the band gap and could favor the formation of the L-phase that has a smaller gap.
Within the density functional perturbation theory, the $Z^{*}$ are defined as \cite{Gonze97}:

\begin{equation}
\label{BEC_eq}
Z^{*A}_{ij}=\frac{1}{q}\frac{\partial F_{i}^A}{\partial E_{j}}\, ,
\end{equation}

with i,j=x,y,z, and where E$_{j}$ is the electric field applied along the j-direction while F$_{i}^A$ is the force that acts on the ion along the i-direction one the atom A.\\

The $Z^{*}$ are reported in Tab. \ref{BEC}. As expected, diagonal $Z^{*}$ are positive for the cation Ca and Ru, while are negative for the oxygen anions. We have obtained large $Z^{*}$ especially for the cation Ca, similar large $Z^{*}$ were observed also in other transition-metal oxides.\cite{PhysRevB.56.983}
The largest $Z^{*}$ diagonal element for the Ru atoms is given by $Z^{*}_{zz}$=1.193 $|e|$ which indicates that under an electric field along the z-direction, a component of the force acting on the Ru atoms is along the z-axis.  The diagonal terms are the largest for both Ca and O atoms, but not for the Ru. 
Indeed, we have found large off-diagonal terms for the Ru atoms up to $Z^{*}_{yz}$=3.121 $|e|$ which is much larger than $Z^{*}_{zz}$. Therefore, under an electric field along the z-direction there is also a component of the force on the Ru atoms acting along the y-axis. Large $Z^{*}$ off-diagonal terms were already observed for highly distorted transition metals perovskites, \cite{offdiagonal_BEC} even if we did not find in the literature cases where off-diagonal $Z^{*}$ terms are larger than diagonal $Z^{*}$ terms as they appear for the Ru.
Due to the symmetry of the system, the Z$^{*}_{ij}$ tensor is not diagonalizable as we can see for the Ru case with spin up:
Z$^{*Ru_{\uparrow}}_{ij}=
\begin{pmatrix}
0.045 &   -0.464 &   -0.119\\
0.412 & \hspace{0.3cm}0.037 & \hspace{0.3cm}3.121\\
0.409 & \hspace{0.3cm}0.377 & \hspace{0.3cm}1.193
\end{pmatrix}$.
A diagonal Z$^{*}_{ij}$ tensor in perovskites can be obtained in absence of octahedral tilts and distortions, however, the undistorted Ca$_2$RuO$_4$ would be metallic as for the Sr$_2$RuO$_4$.
\\

\subsection{Properties of the bulk without inversion symmetry}

We also calculate the piezoelectric tensor, which is zero for centrosymmetric crystal structures \cite{Zou15}. Therefore, in order to obtain a piezoelectric tensor different from zero in systems like Ca$_2$RuO$_4$ we have to break the inversion symmetry \cite{Kuwata80,Khanbabaee16,Park22}.
We have shifted the Ru atoms along the z-axis to break the inversion symmetry, as shown in Fig. \ref{crystal_structure}a), and we have calculated the components of the tensor. 
The Ru atoms have been moved by 0.023, 0.046 and 0.069 {\AA} along the positive direction of the z-axis and we studied how the piezoelectric components  vary at different values of the positions of the Ru atoms, for both clamped and relaxed contributions.
The piezoelectric tensor is defined as \cite{Catti03}: 

\begin{equation}
\label{piezo}
\epsilon^{0}_{ij}=-\frac{\partial \sigma_{i}}{\partial E_{j}}\, ,
\end{equation}

where i=xx,yy,zz,xy,yz,zx and j=x,y,z. We can also use the notation i=1,2,3,4,5,6 and j=1,2,3.

The results are reported in Fig. \ref{Piezoelectric tensor}. We report both the ion-clamped (panels a and b) and the relaxed contributions (panels c and d), where the latter ones are the contributions that include ionic relaxation.
With diagonal and off-diagonal we refer to the components of the stress tensor, namely the diagonal components are i=xx,yy,zz (i=1,2,3) while off-diagonal are those with i=xy,yz,zx (i=4,5,6).
We can see from the figure that, both in the cases of ion-clamped and relaxed contributions, the diagonal terms are one order of magnitude greater than the off-diagonal elements. 
The diagonal terms are zero in the case of centrosymmetric crystal structure, while they become different from zero when we break the inversion symmetry and they are almost constant at different values of the displacement of the Ru atoms, while the off-diagonal terms show a linear behaviour in the case of ion-clamped contributions while a more complex trend in the case of the relaxed contributions.
From these results, we can state that in case of breaking of the inversion symmetry due for example to the presence of interfaces or electric fields the compound can show piezoelectric features.
The positive value of the diagonal components means that the system would increase its volume under an external static electric field.

\subsection{Properties of the slab at large electric fields}

\begin{figure}[h]
	\centering
	\includegraphics[width=9.3cm,angle=0]{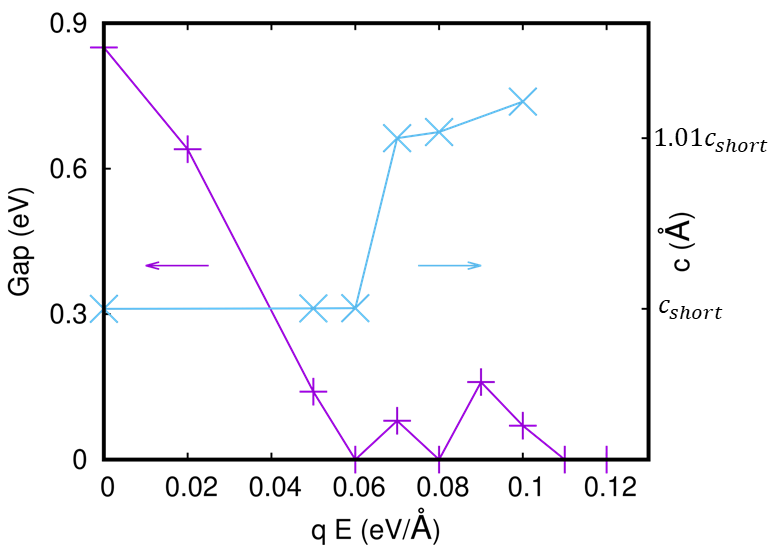}
	\caption{Gap and equilibrium c-axis as a function of the electric field in the slab for the S-Pbca phase.}
	\label{Ca2RuO4_SLAB_c}
\end{figure}

We have analysed a four unit cells slab with a vacuum of 20 {\AA} as shown in Fig. \ref{crystal_structure}b) for both the S- and L-Pbca phase. We have compressed and elongated the lattice constant c and we have studied the slab without and with the application of an external electric field.
In absence of the electric field, the results show a shift of the equilibrium c-axis to lower values respect to the bulk for both the L- and S-Pbca phases.
Then, we studied the slab under the application of an external electric field.  
In Fig. \ref{Ca2RuO4_SLAB_c} the gap at a fixed value of the c axis, namely c$_{short}$, and the equilibrium c axis for the S-Pbca phase as a function of the electric field are reported.
We can see that great values of the electric field close the gap and bring the system to a metallic state.
The gap gets closed at qE=0.06 eV/{\AA}=6$\times$10$^6$ eV/cm and a new gap appears twice, then the electric field definitively closes the gap at 0.11 eV/{\AA}.

\begin{figure}[h]
	\centering
	\includegraphics[width=6.0cm,angle=270]{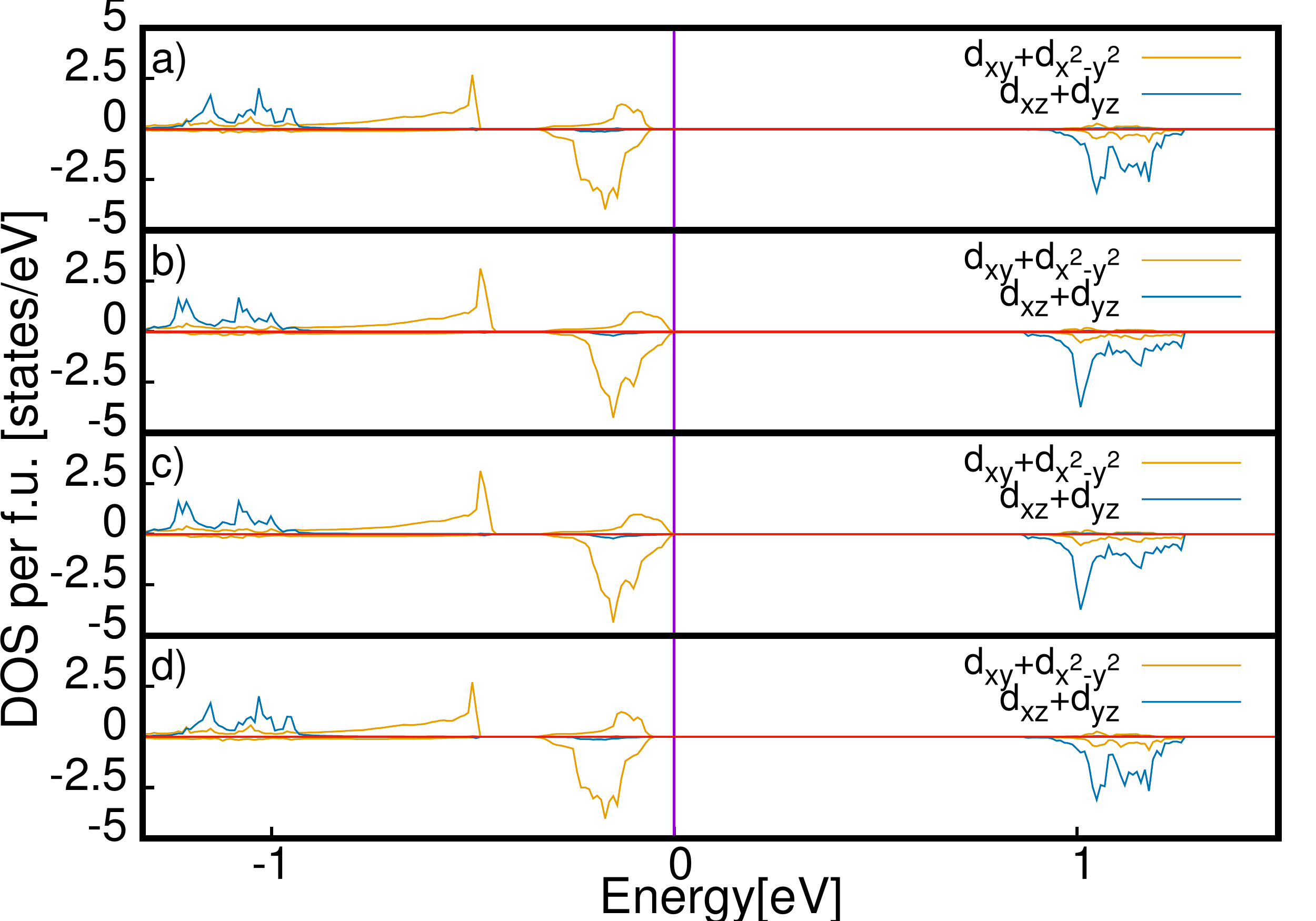}
	\caption{Local density of states (LDOS) for the $d$-orbitals of the Ru atoms of the four layers at E=0 eV for the S-Pbca phase in a range between -5 and 5 eV. The LDOS related to d$_{xz}$ and d$_{yz}$ orbitals is divided by 3 for a better visualization. 
	We report the LDOS for the unit cells u1, u2, u3 and u4 in panels a), b) c) and d), respectively.}
	\label{Ca2RuO4_DOS_E0}
\end{figure}

\begin{figure}[h]
	\centering
	\includegraphics[width=6.0cm,angle=270]{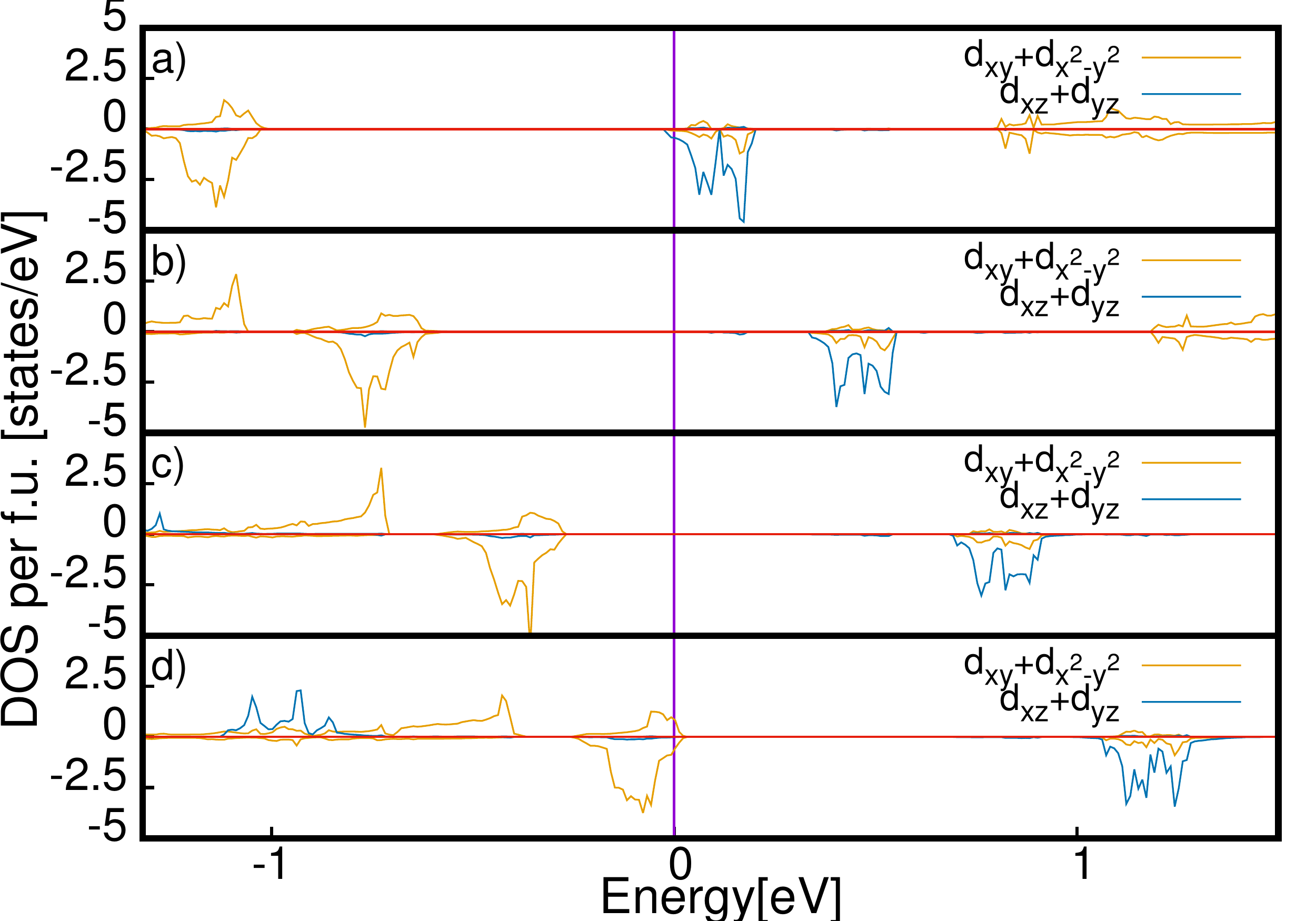}
	\caption{Local density of states for the $d$-orbitals of the Ru atoms of the four layers at qE=0.06 eV/{\AA} for the S-Pbca phase in a range between -5 and 5 eV. The LDOS related to d$_{xz}$ and d$_{yz}$ orbitals is divided by 3 for a better visualization. We report the LDOS for the unit cells u1, u2, u3 and u4 in panels a), b) c) and d), respectively.}
	\label{Ca2RuO4_DOS_E006}
\end{figure}

There is a phase transition in the equilibrium c-axis when the gap gets closed at qE=0.06 eV/{\AA}. We have used a parabolic fit of the values of the equilibrium c-axis as a function of the applied electric field and we have found that the values of c always increase as the values of the applied field increase.
We also report the local density of states projected on the $d$-orbitals of the Ru atoms of the four surfaces. 
We show the DOS at two fields, qE=0 and qE=0.06 eV/{\AA}, in  Fig. \ref{Ca2RuO4_DOS_E0} and \ref{Ca2RuO4_DOS_E006}, respectively. In the first case, the system is insulating and in the second it is metallic.
By looking at the local density of states, we can state that the electric field at qE=0.06 eV/{\AA} closes the gap on the surfaces while the rest of the sample is insulating. The system has metallic layers in the same fashion as the two-dimensional electron gas.

\begin{figure}[h]
	\centering
	\includegraphics[width=6.3cm,angle=270]{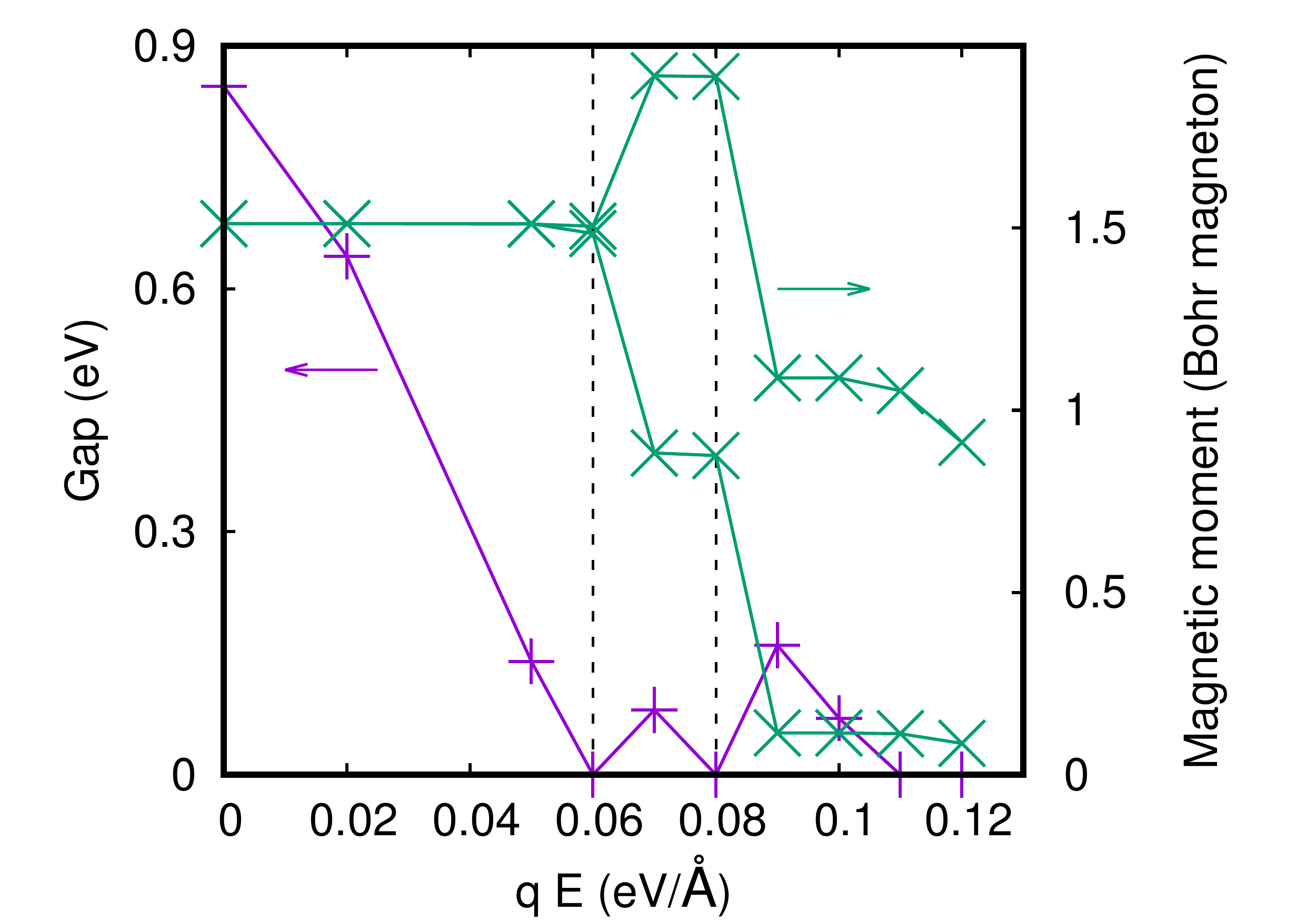}
	\caption{Gap and magnetic moments of the Ru atoms of the two surfaces as a function of the electric field in the slab for the S-Pbca phase.}
	\label{Ca2RuO4_SLAB_m}
\end{figure}

\begin{figure}[h]
	\centering
	\includegraphics[width=6.3cm,angle=270]{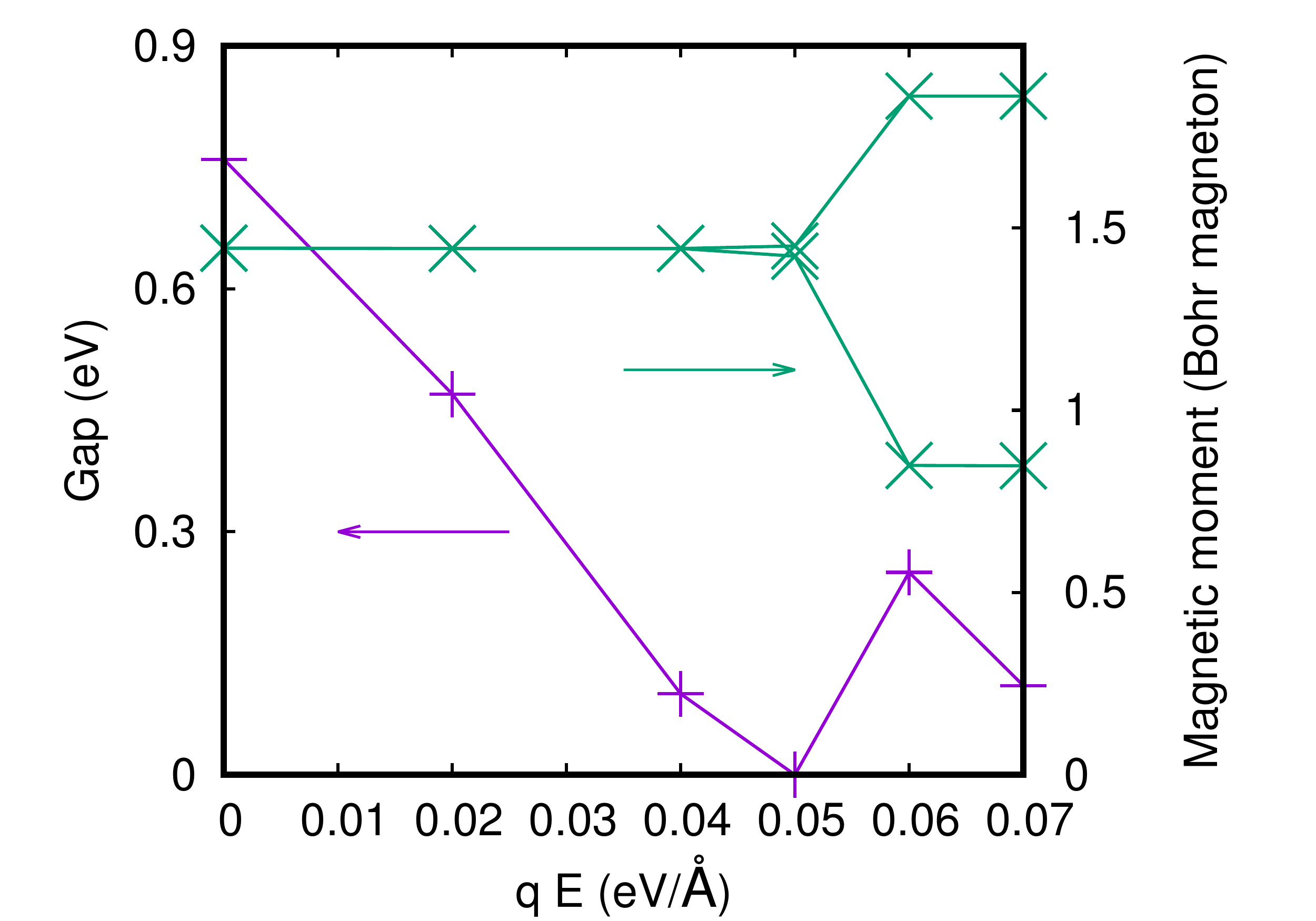}
	\caption{Gap and magnetic moments of the Ru atoms of the two surfaces as a function of the electric field in the slab for the L-Pbca phase.}
	\label{Long_Ca2RuO4_SLAB_m}
\end{figure}

In Fig. \ref{Ca2RuO4_SLAB_m} the gap and the magnetic moments of the Ru atoms of the two surfaces at a fixed value of the c axis (c$_{short}$) for the S-Pbca phase are reported.
We can identify two phase transitions in the magnetic moments.
The magnetic moments of the two surfaces have opposite behaviour starting from the point where the gap is closed the first time, namely at qE=0.06 eV/{\AA}, one magnetic moment decreases and another increases its value. Then, starting from qE=0.08 eV/{\AA} both magnetic moments decrease, but only one Ru surface gets demagnetized. 
For the L-Pbca phase, we get very similar results but the closing of the gap is shifted to lower values of the electric field, because the gap is smaller in the L-Pbca phase. The magnetic moments of the Ru and the gap for the L-Pbca phase as a function of the electric field at a fixed value of the c axis (c$_{long}$) are reported in Fig. \ref{Long_Ca2RuO4_SLAB_m}.

The competition between these different metallic and 
insulating phases in CRO starts from electric fields around 0.06 eV/{\AA} in the S-phase. At the electric fields studied in the bulk, we are very far from the insulating-metal transition.
As a consequence of the volume changes under electric field, there will be also changes in the octahedral rotations\cite{Kyung2021} that we do not analyze tail here.
\medskip

\section{Conclusions}

We have studied the structural, electronic and magnetic properties of Ca$_2$RuO$_4$ under the application of an external electric field using first-principles calculations. We have analyzed the bulk at low electric fields and a slab of four unit cells at large fields.  In the region of small electric fields we have studied the Ca$_2$RuO$_4$ bulk and we have found that the system is insulating. We have calculated the Born effective charges and the piezoelectric tensor. 
To simulate the presence of interfaces and surfaces, we have broken the inversion symmetry off-centering the Ru atom to obtain the piezoelectric tensor different from zero. The positive value of the diagonal elements of the piezoelectric tensor reveals that the system tends to elongate under applied electric field.\\ 
Finally, we have studied a four unit cells slab up to large electric fields and we have found an insulator-metal transition occurring in the system at large electric fields. If we look at the density of states when the gap is closed, we can see that this happens in two of the four unit cells considered, therefore we are in a two-dimensional electron gas phase. The competition between these different metallic and insulating phases in CRO starts from electric fields around 0.06 eV/{\AA} in the S-phase. Regarding the magnetic properties, we have found two phase transitions in the magnetic moments.  For the L-Pbca phase we get very similar results but the closing of the gap is shifted to lower values of the electric field.\\ 
Even considering the modeling limitations due to the absence of many-body and dynamical effects, in all simulated cases, the static electric field increases the lattice constant c and reduces the band gap of Ca$_2$RuO$_4$. This plays a role in the competition between the L-Pbca phase and the S-Pbca phase especially close to the transition temperature T$_{MI}$.

\acknowledgments{The work is supported by the Foundation for Polish Science through the International
Research Agendas program co-financed by the European Union within the Smart Growth Operational
Programme. We thank M. Cuoco and F. Forte for useful discussions. 
We acknowledge the access to the computing facilities of the Interdisciplinary
Center of Modeling at the University of Warsaw, Grants  G75-10, G84-0 and GB84-1.
We acknowledge the CINECA award under the ISCRA initiatives  IsC93 "RATIO" and IsC99 "SILENTS" grant, for the availability of high-performance computing resources and support.}

\bibliography{Ca2RuO4}

\begin{thebibliography}{54}%
\makeatletter
\providecommand \@ifxundefined [1]{%
 \@ifx{#1\undefined}
}%
\providecommand \@ifnum [1]{%
 \ifnum #1\expandafter \@firstoftwo
 \else \expandafter \@secondoftwo
 \fi
}%
\providecommand \@ifx [1]{%
 \ifx #1\expandafter \@firstoftwo
 \else \expandafter \@secondoftwo
 \fi
}%
\providecommand \natexlab [1]{#1}%
\providecommand \enquote  [1]{``#1''}%
\providecommand \bibnamefont  [1]{#1}%
\providecommand \bibfnamefont [1]{#1}%
\providecommand \citenamefont [1]{#1}%
\providecommand \href@noop [0]{\@secondoftwo}%
\providecommand \href [0]{\begingroup \@sanitize@url \@href}%
\providecommand \@href[1]{\@@startlink{#1}\@@href}%
\providecommand \@@href[1]{\endgroup#1\@@endlink}%
\providecommand \@sanitize@url [0]{\catcode `\\12\catcode `\$12\catcode
  `\&12\catcode `\#12\catcode `\^12\catcode `\_12\catcode `\%12\relax}%
\providecommand \@@startlink[1]{}%
\providecommand \@@endlink[0]{}%
\providecommand \url  [0]{\begingroup\@sanitize@url \@url }%
\providecommand \@url [1]{\endgroup\@href {#1}{\urlprefix }}%
\providecommand \urlprefix  [0]{URL }%
\providecommand \Eprint [0]{\href }%
\providecommand \doibase [0]{http://dx.doi.org/}%
\providecommand \selectlanguage [0]{\@gobble}%
\providecommand \bibinfo  [0]{\@secondoftwo}%
\providecommand \bibfield  [0]{\@secondoftwo}%
\providecommand \translation [1]{[#1]}%
\providecommand \BibitemOpen [0]{}%
\providecommand \bibitemStop [0]{}%
\providecommand \bibitemNoStop [0]{.\EOS\space}%
\providecommand \EOS [0]{\spacefactor3000\relax}%
\providecommand \BibitemShut  [1]{\csname bibitem#1\endcsname}%
\let\auto@bib@innerbib\@empty
\bibitem [{\citenamefont {Imada}\ \emph {et~al.}(1998)\citenamefont {Imada},
  \citenamefont {Fujimori},\ and\ \citenamefont {Tokura}}]{Imada98}%
  \BibitemOpen
  \bibfield  {author} {\bibinfo {author} {\bibfnamefont {M.}~\bibnamefont
  {Imada}}, \bibinfo {author} {\bibfnamefont {A.}~\bibnamefont {Fujimori}}, \
  and\ \bibinfo {author} {\bibfnamefont {Y.}~\bibnamefont {Tokura}},\ }\href
  {\doibase 10.1103/RevModPhys.70.1039} {\bibfield  {journal} {\bibinfo
  {journal} {Rev. Mod. Phys.}\ }\textbf {\bibinfo {volume} {70}},\ \bibinfo
  {pages} {1039} (\bibinfo {year} {1998})}\BibitemShut {NoStop}%
\bibitem [{\citenamefont {Meijer}(2008)}]{Meijer08}%
  \BibitemOpen
  \bibfield  {author} {\bibinfo {author} {\bibfnamefont {G.~I.}\ \bibnamefont
  {Meijer}},\ }\href {\doibase 10.1126/science.1153909} {\bibfield  {journal}
  {\bibinfo  {journal} {Science}\ }\textbf {\bibinfo {volume} {319}} (\bibinfo
  {year} {2008}),\ 10.1126/science.1153909}\BibitemShut {NoStop}%
\bibitem [{\citenamefont {Waser}\ and\ \citenamefont {Aono}(2007)}]{Waser07}%
  \BibitemOpen
  \bibfield  {author} {\bibinfo {author} {\bibfnamefont {R.}~\bibnamefont
  {Waser}}\ and\ \bibinfo {author} {\bibfnamefont {M.}~\bibnamefont {Aono}},\
  }\href {\doibase 10.1038/nmat2023} {\bibfield  {journal} {\bibinfo  {journal}
  {Nature Materials}\ }\textbf {\bibinfo {volume} {6}},\ \bibinfo {pages} {833}
  (\bibinfo {year} {2007})}\BibitemShut {NoStop}%
\bibitem [{\citenamefont {Rozenberg}\ \emph {et~al.}(2004)\citenamefont
  {Rozenberg}, \citenamefont {Inoue},\ and\ \citenamefont
  {S\'anchez}}]{Rozenberg04}%
  \BibitemOpen
  \bibfield  {author} {\bibinfo {author} {\bibfnamefont {M.~J.}\ \bibnamefont
  {Rozenberg}}, \bibinfo {author} {\bibfnamefont {I.~H.}\ \bibnamefont
  {Inoue}}, \ and\ \bibinfo {author} {\bibfnamefont {M.~J.}\ \bibnamefont
  {S\'anchez}},\ }\href {\doibase 10.1103/PhysRevLett.92.178302} {\bibfield
  {journal} {\bibinfo  {journal} {Phys. Rev. Lett.}\ }\textbf {\bibinfo
  {volume} {92}},\ \bibinfo {pages} {178302} (\bibinfo {year}
  {2004})}\BibitemShut {NoStop}%
\bibitem [{\citenamefont {Inoue}\ and\ \citenamefont
  {Rozenberg}(2008)}]{Inoue08}%
  \BibitemOpen
  \bibfield  {author} {\bibinfo {author} {\bibfnamefont {I.~H.}\ \bibnamefont
  {Inoue}}\ and\ \bibinfo {author} {\bibfnamefont {M.~J.}\ \bibnamefont
  {Rozenberg}},\ }\href {\doibase https://doi.org/10.1002/adfm.200800558}
  {\bibfield  {journal} {\bibinfo  {journal} {Advanced Functional Materials}\
  }\textbf {\bibinfo {volume} {18}},\ \bibinfo {pages} {2289} (\bibinfo {year}
  {2008})}\BibitemShut {NoStop}%
\bibitem [{\citenamefont {{van Thiel}}\ \emph {et~al.}(2020)\citenamefont {{van
  Thiel}}, \citenamefont {Fowlie}, \citenamefont {Autieri}, \citenamefont
  {Manca}, \citenamefont {{\v{S}}i{\v{s}}kins}, \citenamefont {Afanasiev},
  \citenamefont {Gariglio},\ and\ \citenamefont
  {Caviglia}}]{vanthiel2020coupling}%
  \BibitemOpen
  \bibfield  {author} {\bibinfo {author} {\bibfnamefont {T.~C.}\ \bibnamefont
  {{van Thiel}}}, \bibinfo {author} {\bibfnamefont {J.}~\bibnamefont {Fowlie}},
  \bibinfo {author} {\bibfnamefont {C.}~\bibnamefont {Autieri}}, \bibinfo
  {author} {\bibfnamefont {N.}~\bibnamefont {Manca}}, \bibinfo {author}
  {\bibfnamefont {M.}~\bibnamefont {{\v{S}}i{\v{s}}kins}}, \bibinfo {author}
  {\bibfnamefont {D.}~\bibnamefont {Afanasiev}}, \bibinfo {author}
  {\bibfnamefont {S.}~\bibnamefont {Gariglio}}, \ and\ \bibinfo {author}
  {\bibfnamefont {A.~D.}\ \bibnamefont {Caviglia}},\ }\href {\doibase
  10.1021/acsmaterialslett.9b00540} {\bibfield  {journal} {\bibinfo  {journal}
  {ACS Materials Lett.}\ }\textbf {\bibinfo {volume} {2}},\ \bibinfo {pages}
  {389} (\bibinfo {year} {2020})},\ \Eprint
  {http://arxiv.org/abs/https://doi.org/10.1021/acsmaterialslett.9b00540}
  {https://doi.org/10.1021/acsmaterialslett.9b00540} \BibitemShut {NoStop}%
\bibitem [{\citenamefont {Autieri}\ \emph {et~al.}(2019)\citenamefont
  {Autieri}, \citenamefont {Barone}, \citenamefont
  {S\l{}awi\ifmmode~\acute{n}\else \'{n}\fi{}ska},\ and\ \citenamefont
  {Picozzi}}]{PhysRevMaterials.3.084416}%
  \BibitemOpen
  \bibfield  {author} {\bibinfo {author} {\bibfnamefont {C.}~\bibnamefont
  {Autieri}}, \bibinfo {author} {\bibfnamefont {P.}~\bibnamefont {Barone}},
  \bibinfo {author} {\bibfnamefont {J.}~\bibnamefont
  {S\l{}awi\ifmmode~\acute{n}\else \'{n}\fi{}ska}}, \ and\ \bibinfo {author}
  {\bibfnamefont {S.}~\bibnamefont {Picozzi}},\ }\href {\doibase
  10.1103/PhysRevMaterials.3.084416} {\bibfield  {journal} {\bibinfo  {journal}
  {Phys. Rev. Materials}\ }\textbf {\bibinfo {volume} {3}},\ \bibinfo {pages}
  {084416} (\bibinfo {year} {2019})}\BibitemShut {NoStop}%
\bibitem [{\citenamefont {Hussain}\ \emph {et~al.}(2022)\citenamefont
  {Hussain}, \citenamefont {Samad}, \citenamefont {Rehman}, \citenamefont
  {Cuono},\ and\ \citenamefont {Autieri}}]{Rashba22}%
  \BibitemOpen
  \bibfield  {author} {\bibinfo {author} {\bibfnamefont {G.}~\bibnamefont
  {Hussain}}, \bibinfo {author} {\bibfnamefont {A.}~\bibnamefont {Samad}},
  \bibinfo {author} {\bibfnamefont {M.~U.}\ \bibnamefont {Rehman}}, \bibinfo
  {author} {\bibfnamefont {G.}~\bibnamefont {Cuono}}, \ and\ \bibinfo {author}
  {\bibfnamefont {C.}~\bibnamefont {Autieri}},\ }\href {\doibase
  10.1016/j.jmmm.2022.169897} {\bibfield  {journal} {\bibinfo  {journal}
  {Journal of Magnetism and Magnetic Materials}\ }\textbf {\bibinfo {volume}
  {563}},\ \bibinfo {pages} {169897} (\bibinfo {year} {2022})}\BibitemShut
  {NoStop}%
\bibitem [{\citenamefont {van Thiel}\ \emph {et~al.}(2021)\citenamefont {van
  Thiel}, \citenamefont {Brzezicki}, \citenamefont {Autieri}, \citenamefont
  {Hortensius}, \citenamefont {Afanasiev}, \citenamefont {Gauquelin},
  \citenamefont {Jannis}, \citenamefont {Janssen}, \citenamefont {Groenendijk},
  \citenamefont {Fatermans}, \citenamefont {Van~Aert}, \citenamefont
  {Verbeeck}, \citenamefont {Cuoco},\ and\ \citenamefont
  {Caviglia}}]{vanThiel2021coupling}%
  \BibitemOpen
  \bibfield  {author} {\bibinfo {author} {\bibfnamefont {T.~C.}\ \bibnamefont
  {van Thiel}}, \bibinfo {author} {\bibfnamefont {W.}~\bibnamefont
  {Brzezicki}}, \bibinfo {author} {\bibfnamefont {C.}~\bibnamefont {Autieri}},
  \bibinfo {author} {\bibfnamefont {J.~R.}\ \bibnamefont {Hortensius}},
  \bibinfo {author} {\bibfnamefont {D.}~\bibnamefont {Afanasiev}}, \bibinfo
  {author} {\bibfnamefont {N.}~\bibnamefont {Gauquelin}}, \bibinfo {author}
  {\bibfnamefont {D.}~\bibnamefont {Jannis}}, \bibinfo {author} {\bibfnamefont
  {N.}~\bibnamefont {Janssen}}, \bibinfo {author} {\bibfnamefont {D.~J.}\
  \bibnamefont {Groenendijk}}, \bibinfo {author} {\bibfnamefont
  {J.}~\bibnamefont {Fatermans}}, \bibinfo {author} {\bibfnamefont
  {S.}~\bibnamefont {Van~Aert}}, \bibinfo {author} {\bibfnamefont
  {J.}~\bibnamefont {Verbeeck}}, \bibinfo {author} {\bibfnamefont
  {M.}~\bibnamefont {Cuoco}}, \ and\ \bibinfo {author} {\bibfnamefont {A.~D.}\
  \bibnamefont {Caviglia}},\ }\href {\doibase 10.1103/PhysRevLett.127.127202}
  {\bibfield  {journal} {\bibinfo  {journal} {Phys. Rev. Lett.}\ }\textbf
  {\bibinfo {volume} {127}},\ \bibinfo {pages} {127202} (\bibinfo {year}
  {2021})}\BibitemShut {NoStop}%
\bibitem [{\citenamefont {Autieri}\ \emph {et~al.}(2014)\citenamefont
  {Autieri}, \citenamefont {Cuoco},\ and\ \citenamefont
  {Noce}}]{PhysRevB.89.075102}%
  \BibitemOpen
  \bibfield  {author} {\bibinfo {author} {\bibfnamefont {C.}~\bibnamefont
  {Autieri}}, \bibinfo {author} {\bibfnamefont {M.}~\bibnamefont {Cuoco}}, \
  and\ \bibinfo {author} {\bibfnamefont {C.}~\bibnamefont {Noce}},\ }\href
  {\doibase 10.1103/PhysRevB.89.075102} {\bibfield  {journal} {\bibinfo
  {journal} {Phys. Rev. B}\ }\textbf {\bibinfo {volume} {89}},\ \bibinfo
  {pages} {075102} (\bibinfo {year} {2014})}\BibitemShut {NoStop}%
\bibitem [{\citenamefont {Autieri}\ \emph {et~al.}(2012)\citenamefont
  {Autieri}, \citenamefont {Cuoco},\ and\ \citenamefont
  {Noce}}]{PhysRevB.85.075126}%
  \BibitemOpen
  \bibfield  {author} {\bibinfo {author} {\bibfnamefont {C.}~\bibnamefont
  {Autieri}}, \bibinfo {author} {\bibfnamefont {M.}~\bibnamefont {Cuoco}}, \
  and\ \bibinfo {author} {\bibfnamefont {C.}~\bibnamefont {Noce}},\ }\href
  {\doibase 10.1103/PhysRevB.85.075126} {\bibfield  {journal} {\bibinfo
  {journal} {Phys. Rev. B}\ }\textbf {\bibinfo {volume} {85}},\ \bibinfo
  {pages} {075126} (\bibinfo {year} {2012})}\BibitemShut {NoStop}%
\bibitem [{\citenamefont {Paul}\ \emph {et~al.}(2014)\citenamefont {Paul},
  \citenamefont {Reitinger}, \citenamefont {Autieri}, \citenamefont {Sanyal},
  \citenamefont {Kreuzpaintner}, \citenamefont {Jutimoosik}, \citenamefont
  {Yimnirun}, \citenamefont {Bern}, \citenamefont {Esquinazi}, \citenamefont
  {Korelis},\ and\ \citenamefont {Böni}}]{Amitesh2014APL}%
  \BibitemOpen
  \bibfield  {author} {\bibinfo {author} {\bibfnamefont {A.}~\bibnamefont
  {Paul}}, \bibinfo {author} {\bibfnamefont {C.}~\bibnamefont {Reitinger}},
  \bibinfo {author} {\bibfnamefont {C.}~\bibnamefont {Autieri}}, \bibinfo
  {author} {\bibfnamefont {B.}~\bibnamefont {Sanyal}}, \bibinfo {author}
  {\bibfnamefont {W.}~\bibnamefont {Kreuzpaintner}}, \bibinfo {author}
  {\bibfnamefont {J.}~\bibnamefont {Jutimoosik}}, \bibinfo {author}
  {\bibfnamefont {R.}~\bibnamefont {Yimnirun}}, \bibinfo {author}
  {\bibfnamefont {F.}~\bibnamefont {Bern}}, \bibinfo {author} {\bibfnamefont
  {P.}~\bibnamefont {Esquinazi}}, \bibinfo {author} {\bibfnamefont
  {P.}~\bibnamefont {Korelis}}, \ and\ \bibinfo {author} {\bibfnamefont
  {P.}~\bibnamefont {Böni}},\ }\href {\doibase 10.1063/1.4885316} {\bibfield
  {journal} {\bibinfo  {journal} {Applied Physics Letters}\ }\textbf {\bibinfo
  {volume} {105}},\ \bibinfo {pages} {022409} (\bibinfo {year} {2014})},\
  \Eprint {http://arxiv.org/abs/https://doi.org/10.1063/1.4885316}
  {https://doi.org/10.1063/1.4885316} \BibitemShut {NoStop}%
\bibitem [{\citenamefont {Autieri}\ and\ \citenamefont
  {Sanyal}(2014)}]{Autieri2014NJP}%
  \BibitemOpen
  \bibfield  {author} {\bibinfo {author} {\bibfnamefont {C.}~\bibnamefont
  {Autieri}}\ and\ \bibinfo {author} {\bibfnamefont {B.}~\bibnamefont
  {Sanyal}},\ }\href {\doibase 10.1088/1367-2630/16/11/113031} {\bibfield
  {journal} {\bibinfo  {journal} {New Journal of Physics}\ }\textbf {\bibinfo
  {volume} {16}},\ \bibinfo {pages} {113031} (\bibinfo {year}
  {2014})}\BibitemShut {NoStop}%
\bibitem [{\citenamefont {Hausmann}\ \emph {et~al.}(2017)\citenamefont
  {Hausmann}, \citenamefont {Ye}, \citenamefont {Aoki}, \citenamefont {Zheng},
  \citenamefont {Stahn}, \citenamefont {Bern}, \citenamefont {Chen},
  \citenamefont {Autieri}, \citenamefont {Sanyal}, \citenamefont {Esquinazi},
  \citenamefont {B{\"o}ni},\ and\ \citenamefont {Paul}}]{Hausmann2017}%
  \BibitemOpen
  \bibfield  {author} {\bibinfo {author} {\bibfnamefont {S.}~\bibnamefont
  {Hausmann}}, \bibinfo {author} {\bibfnamefont {J.}~\bibnamefont {Ye}},
  \bibinfo {author} {\bibfnamefont {T.}~\bibnamefont {Aoki}}, \bibinfo {author}
  {\bibfnamefont {J.-G.}\ \bibnamefont {Zheng}}, \bibinfo {author}
  {\bibfnamefont {J.}~\bibnamefont {Stahn}}, \bibinfo {author} {\bibfnamefont
  {F.}~\bibnamefont {Bern}}, \bibinfo {author} {\bibfnamefont {B.}~\bibnamefont
  {Chen}}, \bibinfo {author} {\bibfnamefont {C.}~\bibnamefont {Autieri}},
  \bibinfo {author} {\bibfnamefont {B.}~\bibnamefont {Sanyal}}, \bibinfo
  {author} {\bibfnamefont {P.~D.}\ \bibnamefont {Esquinazi}}, \bibinfo {author}
  {\bibfnamefont {P.}~\bibnamefont {B{\"o}ni}}, \ and\ \bibinfo {author}
  {\bibfnamefont {A.}~\bibnamefont {Paul}},\ }\href {\doibase
  10.1038/s41598-017-10194-4} {\bibfield  {journal} {\bibinfo  {journal}
  {Scientific Reports}\ }\textbf {\bibinfo {volume} {7}},\ \bibinfo {pages}
  {10734} (\bibinfo {year} {2017})}\BibitemShut {NoStop}%
\bibitem [{\citenamefont {Nakatsuji}\ and\ \citenamefont
  {Maeno}(2000)}]{Nakatsuji00}%
  \BibitemOpen
  \bibfield  {author} {\bibinfo {author} {\bibfnamefont {S.}~\bibnamefont
  {Nakatsuji}}\ and\ \bibinfo {author} {\bibfnamefont {Y.}~\bibnamefont
  {Maeno}},\ }\href {\doibase 10.1103/PhysRevLett.84.2666} {\bibfield
  {journal} {\bibinfo  {journal} {Phys. Rev. Lett.}\ }\textbf {\bibinfo
  {volume} {84}},\ \bibinfo {pages} {2666} (\bibinfo {year}
  {2000})}\BibitemShut {NoStop}%
\bibitem [{\citenamefont {Cuoco}\ \emph {et~al.}(2006)\citenamefont {Cuoco},
  \citenamefont {Forte},\ and\ \citenamefont {Noce}}]{Cuoco06}%
  \BibitemOpen
  \bibfield  {author} {\bibinfo {author} {\bibfnamefont {M.}~\bibnamefont
  {Cuoco}}, \bibinfo {author} {\bibfnamefont {F.}~\bibnamefont {Forte}}, \ and\
  \bibinfo {author} {\bibfnamefont {C.}~\bibnamefont {Noce}},\ }\href {\doibase
  10.1103/PhysRevB.74.195124} {\bibfield  {journal} {\bibinfo  {journal} {Phys.
  Rev. B}\ }\textbf {\bibinfo {volume} {74}},\ \bibinfo {pages} {195124}
  (\bibinfo {year} {2006})}\BibitemShut {NoStop}%
\bibitem [{\citenamefont {Forte}\ \emph {et~al.}(2010)\citenamefont {Forte},
  \citenamefont {Cuoco},\ and\ \citenamefont {Noce}}]{Forte10}%
  \BibitemOpen
  \bibfield  {author} {\bibinfo {author} {\bibfnamefont {F.}~\bibnamefont
  {Forte}}, \bibinfo {author} {\bibfnamefont {M.}~\bibnamefont {Cuoco}}, \ and\
  \bibinfo {author} {\bibfnamefont {C.}~\bibnamefont {Noce}},\ }\href {\doibase
  10.1103/PhysRevB.82.155104} {\bibfield  {journal} {\bibinfo  {journal} {Phys.
  Rev. B}\ }\textbf {\bibinfo {volume} {82}},\ \bibinfo {pages} {155104}
  (\bibinfo {year} {2010})}\BibitemShut {NoStop}%
\bibitem [{\citenamefont {Pincini}\ \emph {et~al.}(2019)\citenamefont
  {Pincini}, \citenamefont {Veiga}, \citenamefont {Dashwood}, \citenamefont
  {Forte}, \citenamefont {Cuoco}, \citenamefont {Perry}, \citenamefont
  {Bencok}, \citenamefont {Boothroyd},\ and\ \citenamefont
  {McMorrow}}]{Pincini19}%
  \BibitemOpen
  \bibfield  {author} {\bibinfo {author} {\bibfnamefont {D.}~\bibnamefont
  {Pincini}}, \bibinfo {author} {\bibfnamefont {L.~S.~I.}\ \bibnamefont
  {Veiga}}, \bibinfo {author} {\bibfnamefont {C.~D.}\ \bibnamefont {Dashwood}},
  \bibinfo {author} {\bibfnamefont {F.}~\bibnamefont {Forte}}, \bibinfo
  {author} {\bibfnamefont {M.}~\bibnamefont {Cuoco}}, \bibinfo {author}
  {\bibfnamefont {R.~S.}\ \bibnamefont {Perry}}, \bibinfo {author}
  {\bibfnamefont {P.}~\bibnamefont {Bencok}}, \bibinfo {author} {\bibfnamefont
  {A.~T.}\ \bibnamefont {Boothroyd}}, \ and\ \bibinfo {author} {\bibfnamefont
  {D.~F.}\ \bibnamefont {McMorrow}},\ }\href {\doibase
  10.1103/PhysRevB.99.075125} {\bibfield  {journal} {\bibinfo  {journal} {Phys.
  Rev. B}\ }\textbf {\bibinfo {volume} {99}},\ \bibinfo {pages} {075125}
  (\bibinfo {year} {2019})}\BibitemShut {NoStop}%
\bibitem [{\citenamefont {Koga}\ \emph {et~al.}(2004)\citenamefont {Koga},
  \citenamefont {Kawakami}, \citenamefont {Rice},\ and\ \citenamefont
  {Sigrist}}]{Koga04}%
  \BibitemOpen
  \bibfield  {author} {\bibinfo {author} {\bibfnamefont {A.}~\bibnamefont
  {Koga}}, \bibinfo {author} {\bibfnamefont {N.}~\bibnamefont {Kawakami}},
  \bibinfo {author} {\bibfnamefont {T.~M.}\ \bibnamefont {Rice}}, \ and\
  \bibinfo {author} {\bibfnamefont {M.}~\bibnamefont {Sigrist}},\ }\href
  {\doibase 10.1103/PhysRevLett.92.216402} {\bibfield  {journal} {\bibinfo
  {journal} {Phys. Rev. Lett.}\ }\textbf {\bibinfo {volume} {92}},\ \bibinfo
  {pages} {216402} (\bibinfo {year} {2004})}\BibitemShut {NoStop}%
\bibitem [{\citenamefont {Das}\ \emph {et~al.}(2018)\citenamefont {Das},
  \citenamefont {Forte}, \citenamefont {Fittipaldi}, \citenamefont {Fatuzzo},
  \citenamefont {Granata}, \citenamefont {Ivashko}, \citenamefont {Horio},
  \citenamefont {Schindler}, \citenamefont {Dantz}, \citenamefont {Tseng},
  \citenamefont {McNally}, \citenamefont {R\o{}nnow}, \citenamefont {Wan},
  \citenamefont {Christensen}, \citenamefont {Pelliciari}, \citenamefont
  {Olalde-Velasco}, \citenamefont {Kikugawa}, \citenamefont {Neupert},
  \citenamefont {Vecchione}, \citenamefont {Schmitt}, \citenamefont {Cuoco},\
  and\ \citenamefont {Chang}}]{Das18}%
  \BibitemOpen
  \bibfield  {author} {\bibinfo {author} {\bibfnamefont {L.}~\bibnamefont
  {Das}}, \bibinfo {author} {\bibfnamefont {F.}~\bibnamefont {Forte}}, \bibinfo
  {author} {\bibfnamefont {R.}~\bibnamefont {Fittipaldi}}, \bibinfo {author}
  {\bibfnamefont {C.~G.}\ \bibnamefont {Fatuzzo}}, \bibinfo {author}
  {\bibfnamefont {V.}~\bibnamefont {Granata}}, \bibinfo {author} {\bibfnamefont
  {O.}~\bibnamefont {Ivashko}}, \bibinfo {author} {\bibfnamefont
  {M.}~\bibnamefont {Horio}}, \bibinfo {author} {\bibfnamefont
  {F.}~\bibnamefont {Schindler}}, \bibinfo {author} {\bibfnamefont
  {M.}~\bibnamefont {Dantz}}, \bibinfo {author} {\bibfnamefont
  {Y.}~\bibnamefont {Tseng}}, \bibinfo {author} {\bibfnamefont {D.~E.}\
  \bibnamefont {McNally}}, \bibinfo {author} {\bibfnamefont {H.~M.}\
  \bibnamefont {R\o{}nnow}}, \bibinfo {author} {\bibfnamefont {W.}~\bibnamefont
  {Wan}}, \bibinfo {author} {\bibfnamefont {N.~B.}\ \bibnamefont
  {Christensen}}, \bibinfo {author} {\bibfnamefont {J.}~\bibnamefont
  {Pelliciari}}, \bibinfo {author} {\bibfnamefont {P.}~\bibnamefont
  {Olalde-Velasco}}, \bibinfo {author} {\bibfnamefont {N.}~\bibnamefont
  {Kikugawa}}, \bibinfo {author} {\bibfnamefont {T.}~\bibnamefont {Neupert}},
  \bibinfo {author} {\bibfnamefont {A.}~\bibnamefont {Vecchione}}, \bibinfo
  {author} {\bibfnamefont {T.}~\bibnamefont {Schmitt}}, \bibinfo {author}
  {\bibfnamefont {M.}~\bibnamefont {Cuoco}}, \ and\ \bibinfo {author}
  {\bibfnamefont {J.}~\bibnamefont {Chang}},\ }\href {\doibase
  10.1103/PhysRevX.8.011048} {\bibfield  {journal} {\bibinfo  {journal} {Phys.
  Rev. X}\ }\textbf {\bibinfo {volume} {8}},\ \bibinfo {pages} {011048}
  (\bibinfo {year} {2018})}\BibitemShut {NoStop}%
\bibitem [{\citenamefont {Alexander}\ \emph {et~al.}(1999)\citenamefont
  {Alexander}, \citenamefont {Cao}, \citenamefont {Dobrosavljevic},
  \citenamefont {McCall}, \citenamefont {Crow}, \citenamefont {Lochner},\ and\
  \citenamefont {Guertin}}]{Alexander99}%
  \BibitemOpen
  \bibfield  {author} {\bibinfo {author} {\bibfnamefont {C.~S.}\ \bibnamefont
  {Alexander}}, \bibinfo {author} {\bibfnamefont {G.}~\bibnamefont {Cao}},
  \bibinfo {author} {\bibfnamefont {V.}~\bibnamefont {Dobrosavljevic}},
  \bibinfo {author} {\bibfnamefont {S.}~\bibnamefont {McCall}}, \bibinfo
  {author} {\bibfnamefont {J.~E.}\ \bibnamefont {Crow}}, \bibinfo {author}
  {\bibfnamefont {E.}~\bibnamefont {Lochner}}, \ and\ \bibinfo {author}
  {\bibfnamefont {R.~P.}\ \bibnamefont {Guertin}},\ }\href {\doibase
  10.1103/PhysRevB.60.R8422} {\bibfield  {journal} {\bibinfo  {journal} {Phys.
  Rev. B}\ }\textbf {\bibinfo {volume} {60}},\ \bibinfo {pages} {R8422}
  (\bibinfo {year} {1999})}\BibitemShut {NoStop}%
\bibitem [{\citenamefont {Gorelov}\ \emph {et~al.}(2010)\citenamefont
  {Gorelov}, \citenamefont {Karolak}, \citenamefont {Wehling}, \citenamefont
  {Lechermann}, \citenamefont {Lichtenstein},\ and\ \citenamefont
  {Pavarini}}]{Gorelov10}%
  \BibitemOpen
  \bibfield  {author} {\bibinfo {author} {\bibfnamefont {E.}~\bibnamefont
  {Gorelov}}, \bibinfo {author} {\bibfnamefont {M.}~\bibnamefont {Karolak}},
  \bibinfo {author} {\bibfnamefont {T.~O.}\ \bibnamefont {Wehling}}, \bibinfo
  {author} {\bibfnamefont {F.}~\bibnamefont {Lechermann}}, \bibinfo {author}
  {\bibfnamefont {A.~I.}\ \bibnamefont {Lichtenstein}}, \ and\ \bibinfo
  {author} {\bibfnamefont {E.}~\bibnamefont {Pavarini}},\ }\href {\doibase
  10.1103/PhysRevLett.104.226401} {\bibfield  {journal} {\bibinfo  {journal}
  {Phys. Rev. Lett.}\ }\textbf {\bibinfo {volume} {104}},\ \bibinfo {pages}
  {226401} (\bibinfo {year} {2010})}\BibitemShut {NoStop}%
\bibitem [{\citenamefont {Zhang}\ and\ \citenamefont
  {Pavarini}(2017)}]{Zhang17}%
  \BibitemOpen
  \bibfield  {author} {\bibinfo {author} {\bibfnamefont {G.}~\bibnamefont
  {Zhang}}\ and\ \bibinfo {author} {\bibfnamefont {E.}~\bibnamefont
  {Pavarini}},\ }\href {\doibase 10.1103/PhysRevB.95.075145} {\bibfield
  {journal} {\bibinfo  {journal} {Phys. Rev. B}\ }\textbf {\bibinfo {volume}
  {95}},\ \bibinfo {pages} {075145} (\bibinfo {year} {2017})}\BibitemShut
  {NoStop}%
\bibitem [{\citenamefont {Okazaki}\ \emph {et~al.}(2013)\citenamefont
  {Okazaki}, \citenamefont {Nishina}, \citenamefont {Yasui}, \citenamefont
  {Nakamura}, \citenamefont {Suzuki},\ and\ \citenamefont
  {Terasaki}}]{Okazaki13}%
  \BibitemOpen
  \bibfield  {author} {\bibinfo {author} {\bibfnamefont {R.}~\bibnamefont
  {Okazaki}}, \bibinfo {author} {\bibfnamefont {Y.}~\bibnamefont {Nishina}},
  \bibinfo {author} {\bibfnamefont {Y.}~\bibnamefont {Yasui}}, \bibinfo
  {author} {\bibfnamefont {F.}~\bibnamefont {Nakamura}}, \bibinfo {author}
  {\bibfnamefont {T.}~\bibnamefont {Suzuki}}, \ and\ \bibinfo {author}
  {\bibfnamefont {I.}~\bibnamefont {Terasaki}},\ }\href {\doibase
  10.7566/JPSJ.82.103702} {\bibfield  {journal} {\bibinfo  {journal} {Journal
  of the Physical Society of Japan}\ }\textbf {\bibinfo {volume} {82}},\
  \bibinfo {pages} {103702} (\bibinfo {year} {2013})},\ \Eprint
  {http://arxiv.org/abs/https://doi.org/10.7566/JPSJ.82.103702}
  {https://doi.org/10.7566/JPSJ.82.103702} \BibitemShut {NoStop}%
\bibitem [{\citenamefont {Porter}\ \emph {et~al.}(2018)\citenamefont {Porter},
  \citenamefont {Granata}, \citenamefont {Forte}, \citenamefont {Di~Matteo},
  \citenamefont {Cuoco}, \citenamefont {Fittipaldi}, \citenamefont
  {Vecchione},\ and\ \citenamefont {Bombardi}}]{Porter18}%
  \BibitemOpen
  \bibfield  {author} {\bibinfo {author} {\bibfnamefont {D.~G.}\ \bibnamefont
  {Porter}}, \bibinfo {author} {\bibfnamefont {V.}~\bibnamefont {Granata}},
  \bibinfo {author} {\bibfnamefont {F.}~\bibnamefont {Forte}}, \bibinfo
  {author} {\bibfnamefont {S.}~\bibnamefont {Di~Matteo}}, \bibinfo {author}
  {\bibfnamefont {M.}~\bibnamefont {Cuoco}}, \bibinfo {author} {\bibfnamefont
  {R.}~\bibnamefont {Fittipaldi}}, \bibinfo {author} {\bibfnamefont
  {A.}~\bibnamefont {Vecchione}}, \ and\ \bibinfo {author} {\bibfnamefont
  {A.}~\bibnamefont {Bombardi}},\ }\href {\doibase 10.1103/PhysRevB.98.125142}
  {\bibfield  {journal} {\bibinfo  {journal} {Phys. Rev. B}\ }\textbf {\bibinfo
  {volume} {98}},\ \bibinfo {pages} {125142} (\bibinfo {year}
  {2018})}\BibitemShut {NoStop}%
\bibitem [{\citenamefont {Nakamura}\ \emph {et~al.}(2013)\citenamefont
  {Nakamura}, \citenamefont {Sakaki}, \citenamefont {Yamanaka}, \citenamefont
  {Tamaru}, \citenamefont {Suzuki},\ and\ \citenamefont {Maeno}}]{Nakamura13}%
  \BibitemOpen
  \bibfield  {author} {\bibinfo {author} {\bibfnamefont {F.}~\bibnamefont
  {Nakamura}}, \bibinfo {author} {\bibfnamefont {M.}~\bibnamefont {Sakaki}},
  \bibinfo {author} {\bibfnamefont {Y.}~\bibnamefont {Yamanaka}}, \bibinfo
  {author} {\bibfnamefont {S.}~\bibnamefont {Tamaru}}, \bibinfo {author}
  {\bibfnamefont {T.}~\bibnamefont {Suzuki}}, \ and\ \bibinfo {author}
  {\bibfnamefont {Y.}~\bibnamefont {Maeno}},\ }\href {\doibase
  10.1038/srep02536} {\bibfield  {journal} {\bibinfo  {journal} {Scientific
  Reports}\ }\textbf {\bibinfo {volume} {3}} (\bibinfo {year} {2013}),\
  10.1038/srep02536}\BibitemShut {NoStop}%
\bibitem [{\citenamefont {Zhang}\ \emph {et~al.}(2019)\citenamefont {Zhang},
  \citenamefont {McLeod}, \citenamefont {Han}, \citenamefont {Chen},
  \citenamefont {Bechtel}, \citenamefont {Yao}, \citenamefont {Gilbert~Corder},
  \citenamefont {Ciavatti}, \citenamefont {Tao}, \citenamefont {Aronson},
  \citenamefont {Carr}, \citenamefont {Martin}, \citenamefont {Sow},
  \citenamefont {Yonezawa}, \citenamefont {Nakamura}, \citenamefont {Terasaki},
  \citenamefont {Basov}, \citenamefont {Millis}, \citenamefont {Maeno},\ and\
  \citenamefont {Liu}}]{Zhang19}%
  \BibitemOpen
  \bibfield  {author} {\bibinfo {author} {\bibfnamefont {J.}~\bibnamefont
  {Zhang}}, \bibinfo {author} {\bibfnamefont {A.~S.}\ \bibnamefont {McLeod}},
  \bibinfo {author} {\bibfnamefont {Q.}~\bibnamefont {Han}}, \bibinfo {author}
  {\bibfnamefont {X.}~\bibnamefont {Chen}}, \bibinfo {author} {\bibfnamefont
  {H.~A.}\ \bibnamefont {Bechtel}}, \bibinfo {author} {\bibfnamefont
  {Z.}~\bibnamefont {Yao}}, \bibinfo {author} {\bibfnamefont {S.~N.}\
  \bibnamefont {Gilbert~Corder}}, \bibinfo {author} {\bibfnamefont
  {T.}~\bibnamefont {Ciavatti}}, \bibinfo {author} {\bibfnamefont {T.~H.}\
  \bibnamefont {Tao}}, \bibinfo {author} {\bibfnamefont {M.}~\bibnamefont
  {Aronson}}, \bibinfo {author} {\bibfnamefont {G.~L.}\ \bibnamefont {Carr}},
  \bibinfo {author} {\bibfnamefont {M.~C.}\ \bibnamefont {Martin}}, \bibinfo
  {author} {\bibfnamefont {C.}~\bibnamefont {Sow}}, \bibinfo {author}
  {\bibfnamefont {S.}~\bibnamefont {Yonezawa}}, \bibinfo {author}
  {\bibfnamefont {F.}~\bibnamefont {Nakamura}}, \bibinfo {author}
  {\bibfnamefont {I.}~\bibnamefont {Terasaki}}, \bibinfo {author}
  {\bibfnamefont {D.~N.}\ \bibnamefont {Basov}}, \bibinfo {author}
  {\bibfnamefont {A.~J.}\ \bibnamefont {Millis}}, \bibinfo {author}
  {\bibfnamefont {Y.}~\bibnamefont {Maeno}}, \ and\ \bibinfo {author}
  {\bibfnamefont {M.}~\bibnamefont {Liu}},\ }\href {\doibase
  10.1103/PhysRevX.9.011032} {\bibfield  {journal} {\bibinfo  {journal} {Phys.
  Rev. X}\ }\textbf {\bibinfo {volume} {9}},\ \bibinfo {pages} {011032}
  (\bibinfo {year} {2019})}\BibitemShut {NoStop}%
\bibitem [{\citenamefont {Cirillo}\ \emph {et~al.}(2019)\citenamefont
  {Cirillo}, \citenamefont {Granata}, \citenamefont {Avallone}, \citenamefont
  {Fittipaldi}, \citenamefont {Attanasio}, \citenamefont {Avella},\ and\
  \citenamefont {Vecchione}}]{Cirillo19}%
  \BibitemOpen
  \bibfield  {author} {\bibinfo {author} {\bibfnamefont {C.}~\bibnamefont
  {Cirillo}}, \bibinfo {author} {\bibfnamefont {V.}~\bibnamefont {Granata}},
  \bibinfo {author} {\bibfnamefont {G.}~\bibnamefont {Avallone}}, \bibinfo
  {author} {\bibfnamefont {R.}~\bibnamefont {Fittipaldi}}, \bibinfo {author}
  {\bibfnamefont {C.}~\bibnamefont {Attanasio}}, \bibinfo {author}
  {\bibfnamefont {A.}~\bibnamefont {Avella}}, \ and\ \bibinfo {author}
  {\bibfnamefont {A.}~\bibnamefont {Vecchione}},\ }\href {\doibase
  10.1103/PhysRevB.100.235142} {\bibfield  {journal} {\bibinfo  {journal}
  {Phys. Rev. B}\ }\textbf {\bibinfo {volume} {100}},\ \bibinfo {pages}
  {235142} (\bibinfo {year} {2019})}\BibitemShut {NoStop}%
\bibitem [{\citenamefont {Mattoni}\ \emph {et~al.}(2020)\citenamefont
  {Mattoni}, \citenamefont {Yonezawa}, \citenamefont {Nakamura},\ and\
  \citenamefont {Maeno}}]{Mattoni20}%
  \BibitemOpen
  \bibfield  {author} {\bibinfo {author} {\bibfnamefont {G.}~\bibnamefont
  {Mattoni}}, \bibinfo {author} {\bibfnamefont {S.}~\bibnamefont {Yonezawa}},
  \bibinfo {author} {\bibfnamefont {F.}~\bibnamefont {Nakamura}}, \ and\
  \bibinfo {author} {\bibfnamefont {Y.}~\bibnamefont {Maeno}},\ }\href
  {\doibase 10.1103/PhysRevMaterials.4.114414} {\bibfield  {journal} {\bibinfo
  {journal} {Phys. Rev. Materials}\ }\textbf {\bibinfo {volume} {4}},\ \bibinfo
  {pages} {114414} (\bibinfo {year} {2020})}\BibitemShut {NoStop}%
\bibitem [{\citenamefont {Gauquelin}\ \emph {et~al.}(2023)\citenamefont
  {Gauquelin}, \citenamefont {Forte}, \citenamefont {Jannis}, \citenamefont
  {Fittipaldi}, \citenamefont {Autieri}, \citenamefont {Cuono}, \citenamefont
  {Granata}, \citenamefont {Lettieri}, \citenamefont {Noce}, \citenamefont
  {Miletto~Granozio}, \citenamefont {Vecchione}, \citenamefont {Verbeeck},\
  and\ \citenamefont {Cuoco}}]{Gauquelin22}%
  \BibitemOpen
  \bibfield  {author} {\bibinfo {author} {\bibfnamefont {N.}~\bibnamefont
  {Gauquelin}}, \bibinfo {author} {\bibfnamefont {F.}~\bibnamefont {Forte}},
  \bibinfo {author} {\bibfnamefont {D.}~\bibnamefont {Jannis}}, \bibinfo
  {author} {\bibfnamefont {R.}~\bibnamefont {Fittipaldi}}, \bibinfo {author}
  {\bibfnamefont {C.}~\bibnamefont {Autieri}}, \bibinfo {author} {\bibfnamefont
  {G.}~\bibnamefont {Cuono}}, \bibinfo {author} {\bibfnamefont
  {V.}~\bibnamefont {Granata}}, \bibinfo {author} {\bibfnamefont
  {M.}~\bibnamefont {Lettieri}}, \bibinfo {author} {\bibfnamefont
  {C.}~\bibnamefont {Noce}}, \bibinfo {author} {\bibfnamefont {F.}~\bibnamefont
  {Miletto~Granozio}}, \bibinfo {author} {\bibfnamefont {A.}~\bibnamefont
  {Vecchione}}, \bibinfo {author} {\bibfnamefont {J.}~\bibnamefont {Verbeeck}},
  \ and\ \bibinfo {author} {\bibfnamefont {M.}~\bibnamefont {Cuoco}},\
  }\href@noop {} {\bibfield  {journal} {\bibinfo  {journal} {Submitted}\ }
  (\bibinfo {year} {2023})}\BibitemShut {NoStop}%
\bibitem [{\citenamefont {Kresse}\ and\ \citenamefont
  {Hafner}(1993)}]{Kresse93}%
  \BibitemOpen
  \bibfield  {author} {\bibinfo {author} {\bibfnamefont {G.}~\bibnamefont
  {Kresse}}\ and\ \bibinfo {author} {\bibfnamefont {J.}~\bibnamefont
  {Hafner}},\ }\href {\doibase 10.1103/PhysRevB.47.558} {\bibfield  {journal}
  {\bibinfo  {journal} {Phys. Rev. B}\ }\textbf {\bibinfo {volume} {47}},\
  \bibinfo {pages} {558} (\bibinfo {year} {1993})}\BibitemShut {NoStop}%
\bibitem [{\citenamefont {Kresse}\ and\ \citenamefont
  {Furthmüller}(1996)}]{Kresse96a}%
  \BibitemOpen
  \bibfield  {author} {\bibinfo {author} {\bibfnamefont {G.}~\bibnamefont
  {Kresse}}\ and\ \bibinfo {author} {\bibfnamefont {J.}~\bibnamefont
  {Furthmüller}},\ }\href {\doibase
  https://doi.org/10.1016/0927-0256(96)00008-0} {\bibfield  {journal} {\bibinfo
   {journal} {Computational Materials Science}\ }\textbf {\bibinfo {volume}
  {6}},\ \bibinfo {pages} {15} (\bibinfo {year} {1996})}\BibitemShut {NoStop}%
\bibitem [{\citenamefont {Kresse}\ and\ \citenamefont
  {Furthm{\"u}ller}(1996)}]{Kresse96b}%
  \BibitemOpen
  \bibfield  {author} {\bibinfo {author} {\bibfnamefont {G.}~\bibnamefont
  {Kresse}}\ and\ \bibinfo {author} {\bibfnamefont {J.}~\bibnamefont
  {Furthm{\"u}ller}},\ }\href@noop {} {\bibfield  {journal} {\bibinfo
  {journal} {Physical Review B}\ }\textbf {\bibinfo {volume} {54}},\ \bibinfo
  {pages} {11169} (\bibinfo {year} {1996})}\BibitemShut {NoStop}%
\bibitem [{\citenamefont {Kresse}\ and\ \citenamefont
  {Joubert}(1999)}]{Kresse99}%
  \BibitemOpen
  \bibfield  {author} {\bibinfo {author} {\bibfnamefont {G.}~\bibnamefont
  {Kresse}}\ and\ \bibinfo {author} {\bibfnamefont {D.}~\bibnamefont
  {Joubert}},\ }\href {\doibase 10.1103/PhysRevB.59.1758} {\bibfield  {journal}
  {\bibinfo  {journal} {Phys. Rev. B}\ }\textbf {\bibinfo {volume} {59}},\
  \bibinfo {pages} {1758} (\bibinfo {year} {1999})}\BibitemShut {NoStop}%
\bibitem [{\citenamefont {Perdew}\ \emph {et~al.}(2008)\citenamefont {Perdew},
  \citenamefont {Ruzsinszky}, \citenamefont {Csonka}, \citenamefont {Vydrov},
  \citenamefont {Scuseria}, \citenamefont {Constantin}, \citenamefont {Zhou},\
  and\ \citenamefont {Burke}}]{Perdew08}%
  \BibitemOpen
  \bibfield  {author} {\bibinfo {author} {\bibfnamefont {J.~P.}\ \bibnamefont
  {Perdew}}, \bibinfo {author} {\bibfnamefont {A.}~\bibnamefont {Ruzsinszky}},
  \bibinfo {author} {\bibfnamefont {G.~I.}\ \bibnamefont {Csonka}}, \bibinfo
  {author} {\bibfnamefont {O.~A.}\ \bibnamefont {Vydrov}}, \bibinfo {author}
  {\bibfnamefont {G.~E.}\ \bibnamefont {Scuseria}}, \bibinfo {author}
  {\bibfnamefont {L.~A.}\ \bibnamefont {Constantin}}, \bibinfo {author}
  {\bibfnamefont {X.}~\bibnamefont {Zhou}}, \ and\ \bibinfo {author}
  {\bibfnamefont {K.}~\bibnamefont {Burke}},\ }\href {\doibase
  10.1103/PhysRevLett.100.136406} {\bibfield  {journal} {\bibinfo  {journal}
  {Phys. Rev. Lett.}\ }\textbf {\bibinfo {volume} {100}},\ \bibinfo {pages}
  {136406} (\bibinfo {year} {2008})}\BibitemShut {NoStop}%
\bibitem [{\citenamefont {Autieri}(2016)}]{Autieri_2016}%
  \BibitemOpen
  \bibfield  {author} {\bibinfo {author} {\bibfnamefont {C.}~\bibnamefont
  {Autieri}},\ }\href {\doibase 10.1088/0953-8984/28/42/426004} {\bibfield
  {journal} {\bibinfo  {journal} {Journal of Physics: Condensed Matter}\
  }\textbf {\bibinfo {volume} {28}},\ \bibinfo {pages} {426004} (\bibinfo
  {year} {2016})}\BibitemShut {NoStop}%
\bibitem [{\citenamefont {Vaugier}\ \emph {et~al.}(2012)\citenamefont
  {Vaugier}, \citenamefont {Jiang},\ and\ \citenamefont
  {Biermann}}]{Vaugier12}%
  \BibitemOpen
  \bibfield  {author} {\bibinfo {author} {\bibfnamefont {L.}~\bibnamefont
  {Vaugier}}, \bibinfo {author} {\bibfnamefont {H.}~\bibnamefont {Jiang}}, \
  and\ \bibinfo {author} {\bibfnamefont {S.}~\bibnamefont {Biermann}},\ }\href
  {\doibase 10.1103/PhysRevB.86.165105} {\bibfield  {journal} {\bibinfo
  {journal} {Phys. Rev. B}\ }\textbf {\bibinfo {volume} {86}},\ \bibinfo
  {pages} {165105} (\bibinfo {year} {2012})}\BibitemShut {NoStop}%
\bibitem [{\citenamefont {Friedt}\ \emph {et~al.}(2001)\citenamefont {Friedt},
  \citenamefont {Braden}, \citenamefont {Andr\'e}, \citenamefont {Adelmann},
  \citenamefont {Nakatsuji},\ and\ \citenamefont {Maeno}}]{Friedt01}%
  \BibitemOpen
  \bibfield  {author} {\bibinfo {author} {\bibfnamefont {O.}~\bibnamefont
  {Friedt}}, \bibinfo {author} {\bibfnamefont {M.}~\bibnamefont {Braden}},
  \bibinfo {author} {\bibfnamefont {G.}~\bibnamefont {Andr\'e}}, \bibinfo
  {author} {\bibfnamefont {P.}~\bibnamefont {Adelmann}}, \bibinfo {author}
  {\bibfnamefont {S.}~\bibnamefont {Nakatsuji}}, \ and\ \bibinfo {author}
  {\bibfnamefont {Y.}~\bibnamefont {Maeno}},\ }\href {\doibase
  10.1103/PhysRevB.63.174432} {\bibfield  {journal} {\bibinfo  {journal} {Phys.
  Rev. B}\ }\textbf {\bibinfo {volume} {63}},\ \bibinfo {pages} {174432}
  (\bibinfo {year} {2001})}\BibitemShut {NoStop}%
\bibitem [{\citenamefont {Resta}(1992)}]{Resta92}%
  \BibitemOpen
  \bibfield  {author} {\bibinfo {author} {\bibfnamefont {R.}~\bibnamefont
  {Resta}},\ }\href {\doibase 10.1080/00150199208016065} {\bibfield  {journal}
  {\bibinfo  {journal} {Ferroelectrics}\ }\textbf {\bibinfo {volume} {136}},\
  \bibinfo {pages} {51} (\bibinfo {year} {1992})},\ \Eprint
  {http://arxiv.org/abs/https://doi.org/10.1080/00150199208016065}
  {https://doi.org/10.1080/00150199208016065} \BibitemShut {NoStop}%
\bibitem [{\citenamefont {King-Smith}\ and\ \citenamefont
  {Vanderbilt}(1993)}]{Smith93}%
  \BibitemOpen
  \bibfield  {author} {\bibinfo {author} {\bibfnamefont {R.~D.}\ \bibnamefont
  {King-Smith}}\ and\ \bibinfo {author} {\bibfnamefont {D.}~\bibnamefont
  {Vanderbilt}},\ }\href {\doibase 10.1103/PhysRevB.47.1651} {\bibfield
  {journal} {\bibinfo  {journal} {Phys. Rev. B}\ }\textbf {\bibinfo {volume}
  {47}},\ \bibinfo {pages} {1651} (\bibinfo {year} {1993})}\BibitemShut
  {NoStop}%
\bibitem [{\citenamefont {Nunes}\ and\ \citenamefont {Gonze}(2001)}]{Nunes01}%
  \BibitemOpen
  \bibfield  {author} {\bibinfo {author} {\bibfnamefont {R.~W.}\ \bibnamefont
  {Nunes}}\ and\ \bibinfo {author} {\bibfnamefont {X.}~\bibnamefont {Gonze}},\
  }\href {\doibase 10.1103/PhysRevB.63.155107} {\bibfield  {journal} {\bibinfo
  {journal} {Phys. Rev. B}\ }\textbf {\bibinfo {volume} {63}},\ \bibinfo
  {pages} {155107} (\bibinfo {year} {2001})}\BibitemShut {NoStop}%
\bibitem [{\citenamefont {Souza}\ \emph {et~al.}(2002)\citenamefont {Souza},
  \citenamefont {\'I\~niguez},\ and\ \citenamefont {Vanderbilt}}]{Souza02}%
  \BibitemOpen
  \bibfield  {author} {\bibinfo {author} {\bibfnamefont {I.}~\bibnamefont
  {Souza}}, \bibinfo {author} {\bibfnamefont {J.}~\bibnamefont {\'I\~niguez}},
  \ and\ \bibinfo {author} {\bibfnamefont {D.}~\bibnamefont {Vanderbilt}},\
  }\href {\doibase 10.1103/PhysRevLett.89.117602} {\bibfield  {journal}
  {\bibinfo  {journal} {Phys. Rev. Lett.}\ }\textbf {\bibinfo {volume} {89}},\
  \bibinfo {pages} {117602} (\bibinfo {year} {2002})}\BibitemShut {NoStop}%
\bibitem [{\citenamefont {Islam}\ \emph {et~al.}(2021)\citenamefont {Islam},
  \citenamefont {Ghosh}, \citenamefont {Autieri}, \citenamefont {Chowdhury},
  \citenamefont {Bansil}, \citenamefont {Agarwal},\ and\ \citenamefont
  {Singh}}]{Islam21}%
  \BibitemOpen
  \bibfield  {author} {\bibinfo {author} {\bibfnamefont {R.}~\bibnamefont
  {Islam}}, \bibinfo {author} {\bibfnamefont {B.}~\bibnamefont {Ghosh}},
  \bibinfo {author} {\bibfnamefont {C.}~\bibnamefont {Autieri}}, \bibinfo
  {author} {\bibfnamefont {S.}~\bibnamefont {Chowdhury}}, \bibinfo {author}
  {\bibfnamefont {A.}~\bibnamefont {Bansil}}, \bibinfo {author} {\bibfnamefont
  {A.}~\bibnamefont {Agarwal}}, \ and\ \bibinfo {author} {\bibfnamefont
  {B.}~\bibnamefont {Singh}},\ }\href {\doibase 10.1103/PhysRevB.104.L201112}
  {\bibfield  {journal} {\bibinfo  {journal} {Phys. Rev. B}\ }\textbf {\bibinfo
  {volume} {104}},\ \bibinfo {pages} {L201112} (\bibinfo {year}
  {2021})}\BibitemShut {NoStop}%
\bibitem [{\citenamefont {Islam}\ \emph {et~al.}(2022)\citenamefont {Islam},
  \citenamefont {Verma}, \citenamefont {Ghosh}, \citenamefont {Muhammad},
  \citenamefont {Bansil}, \citenamefont {Autieri},\ and\ \citenamefont
  {Singh}}]{Islam22}%
  \BibitemOpen
  \bibfield  {author} {\bibinfo {author} {\bibfnamefont {R.}~\bibnamefont
  {Islam}}, \bibinfo {author} {\bibfnamefont {R.}~\bibnamefont {Verma}},
  \bibinfo {author} {\bibfnamefont {B.}~\bibnamefont {Ghosh}}, \bibinfo
  {author} {\bibfnamefont {Z.}~\bibnamefont {Muhammad}}, \bibinfo {author}
  {\bibfnamefont {A.}~\bibnamefont {Bansil}}, \bibinfo {author} {\bibfnamefont
  {C.}~\bibnamefont {Autieri}}, \ and\ \bibinfo {author} {\bibfnamefont
  {B.}~\bibnamefont {Singh}},\ }\href {\doibase 10.48550/ARXIV.2207.08407}
  {\enquote {\bibinfo {title} {Switchable large-gap quantum spin hall state in
  two-dimensional msi$_2$z$_4$ materials class},}\ } (\bibinfo {year}
  {2022})\BibitemShut {NoStop}%
\bibitem [{\citenamefont {Neugebauer}\ and\ \citenamefont
  {Scheffler}(1992)}]{Neugebauer92}%
  \BibitemOpen
  \bibfield  {author} {\bibinfo {author} {\bibfnamefont {J.}~\bibnamefont
  {Neugebauer}}\ and\ \bibinfo {author} {\bibfnamefont {M.}~\bibnamefont
  {Scheffler}},\ }\href {\doibase 10.1103/PhysRevB.46.16067} {\bibfield
  {journal} {\bibinfo  {journal} {Phys. Rev. B}\ }\textbf {\bibinfo {volume}
  {46}},\ \bibinfo {pages} {16067} (\bibinfo {year} {1992})}\BibitemShut
  {NoStop}%
\bibitem [{\citenamefont {Gonze}\ and\ \citenamefont {Lee}(1997)}]{Gonze97}%
  \BibitemOpen
  \bibfield  {author} {\bibinfo {author} {\bibfnamefont {X.}~\bibnamefont
  {Gonze}}\ and\ \bibinfo {author} {\bibfnamefont {C.}~\bibnamefont {Lee}},\
  }\href {\doibase 10.1103/PhysRevB.55.10355} {\bibfield  {journal} {\bibinfo
  {journal} {Phys. Rev. B}\ }\textbf {\bibinfo {volume} {55}},\ \bibinfo
  {pages} {10355} (\bibinfo {year} {1997})}\BibitemShut {NoStop}%
\bibitem [{\citenamefont {Detraux}\ \emph {et~al.}(1997)\citenamefont
  {Detraux}, \citenamefont {Ghosez},\ and\ \citenamefont
  {Gonze}}]{PhysRevB.56.983}%
  \BibitemOpen
  \bibfield  {author} {\bibinfo {author} {\bibfnamefont {F.}~\bibnamefont
  {Detraux}}, \bibinfo {author} {\bibfnamefont {P.}~\bibnamefont {Ghosez}}, \
  and\ \bibinfo {author} {\bibfnamefont {X.}~\bibnamefont {Gonze}},\ }\href
  {\doibase 10.1103/PhysRevB.56.983} {\bibfield  {journal} {\bibinfo  {journal}
  {Phys. Rev. B}\ }\textbf {\bibinfo {volume} {56}},\ \bibinfo {pages} {983}
  (\bibinfo {year} {1997})}\BibitemShut {NoStop}%
\bibitem [{\citenamefont {Ganga~Prasad}\ \emph {et~al.}(2016)\citenamefont
  {Ganga~Prasad}, \citenamefont {Niranjan}, \citenamefont {Asthana},\ and\
  \citenamefont {Karthikeyan}}]{offdiagonal_BEC}%
  \BibitemOpen
  \bibfield  {author} {\bibinfo {author} {\bibfnamefont {K.}~\bibnamefont
  {Ganga~Prasad}}, \bibinfo {author} {\bibfnamefont {M.~K.}\ \bibnamefont
  {Niranjan}}, \bibinfo {author} {\bibfnamefont {S.}~\bibnamefont {Asthana}}, \
  and\ \bibinfo {author} {\bibfnamefont {R.}~\bibnamefont {Karthikeyan}},\
  }\href {\doibase https://doi.org/10.1111/jace.13929} {\bibfield  {journal}
  {\bibinfo  {journal} {Journal of the American Ceramic Society}\ }\textbf
  {\bibinfo {volume} {99}},\ \bibinfo {pages} {332} (\bibinfo {year}
  {2016})}\BibitemShut {NoStop}%
\bibitem [{\citenamefont {Zou}\ \emph {et~al.}(2013)\citenamefont {Zou},
  \citenamefont {Tang},\ and\ \citenamefont {Pan}}]{Zou15}%
  \BibitemOpen
  \bibfield  {author} {\bibinfo {author} {\bibfnamefont {W.-N.}\ \bibnamefont
  {Zou}}, \bibinfo {author} {\bibfnamefont {C.-X.}\ \bibnamefont {Tang}}, \
  and\ \bibinfo {author} {\bibfnamefont {E.}~\bibnamefont {Pan}},\ }\href
  {\doibase 10.1098/rspa.2012.0755} {\bibfield  {journal} {\bibinfo  {journal}
  {Proceedings of the Royal Society A: Mathematical, Physical and Engineering
  Sciences}\ }\textbf {\bibinfo {volume} {469}},\ \bibinfo {pages} {20120755}
  (\bibinfo {year} {2013})}\BibitemShut {NoStop}%
\bibitem [{\citenamefont {Kuwata}\ \emph {et~al.}(1980)\citenamefont {Kuwata},
  \citenamefont {Uchino},\ and\ \citenamefont {Nomura}}]{Kuwata80}%
  \BibitemOpen
  \bibfield  {author} {\bibinfo {author} {\bibfnamefont {J.}~\bibnamefont
  {Kuwata}}, \bibinfo {author} {\bibfnamefont {K.}~\bibnamefont {Uchino}}, \
  and\ \bibinfo {author} {\bibfnamefont {S.}~\bibnamefont {Nomura}},\ }\href
  {\doibase 10.1143/JJAP.19.2099} {\bibfield  {journal} {\bibinfo  {journal}
  {Japanese Journal of Applied Physics}\ }\textbf {\bibinfo {volume} {19}},\
  \bibinfo {pages} {2099} (\bibinfo {year} {1980})}\BibitemShut {NoStop}%
\bibitem [{\citenamefont {Khanbabaee}\ \emph {et~al.}(2016)\citenamefont
  {Khanbabaee}, \citenamefont {Mehner}, \citenamefont {Richter}, \citenamefont
  {Hanzig}, \citenamefont {Zschornak}, \citenamefont {Pietsch}, \citenamefont
  {Stöcker}, \citenamefont {Leisegang}, \citenamefont {Meyer},\ and\
  \citenamefont {Gorfman}}]{Khanbabaee16}%
  \BibitemOpen
  \bibfield  {author} {\bibinfo {author} {\bibfnamefont {B.}~\bibnamefont
  {Khanbabaee}}, \bibinfo {author} {\bibfnamefont {E.}~\bibnamefont {Mehner}},
  \bibinfo {author} {\bibfnamefont {C.}~\bibnamefont {Richter}}, \bibinfo
  {author} {\bibfnamefont {J.}~\bibnamefont {Hanzig}}, \bibinfo {author}
  {\bibfnamefont {M.}~\bibnamefont {Zschornak}}, \bibinfo {author}
  {\bibfnamefont {U.}~\bibnamefont {Pietsch}}, \bibinfo {author} {\bibfnamefont
  {H.}~\bibnamefont {Stöcker}}, \bibinfo {author} {\bibfnamefont
  {T.}~\bibnamefont {Leisegang}}, \bibinfo {author} {\bibfnamefont {D.~C.}\
  \bibnamefont {Meyer}}, \ and\ \bibinfo {author} {\bibfnamefont
  {S.}~\bibnamefont {Gorfman}},\ }\href {\doibase 10.1063/1.4966892} {\bibfield
   {journal} {\bibinfo  {journal} {Applied Physics Letters}\ }\textbf {\bibinfo
  {volume} {109}},\ \bibinfo {pages} {222901} (\bibinfo {year} {2016})},\
  \Eprint {http://arxiv.org/abs/https://doi.org/10.1063/1.4966892}
  {https://doi.org/10.1063/1.4966892} \BibitemShut {NoStop}%
\bibitem [{\citenamefont {Park}\ and\ \citenamefont {et~al.}(2022)}]{Park22}%
  \BibitemOpen
  \bibfield  {author} {\bibinfo {author} {\bibfnamefont {D.}~\bibnamefont
  {Park}}\ and\ \bibinfo {author} {\bibnamefont {et~al.}},\ }\href {\doibase
  10.1126/science.abm7497} {\bibfield  {journal} {\bibinfo  {journal}
  {Science}\ }\textbf {\bibinfo {volume} {375}},\ \bibinfo {pages} {653}
  (\bibinfo {year} {2022})}\BibitemShut {NoStop}%
\bibitem [{\citenamefont {Catti}\ \emph {et~al.}(2003)\citenamefont {Catti},
  \citenamefont {Noel},\ and\ \citenamefont {Dovesi}}]{Catti03}%
  \BibitemOpen
  \bibfield  {author} {\bibinfo {author} {\bibfnamefont {M.}~\bibnamefont
  {Catti}}, \bibinfo {author} {\bibfnamefont {Y.}~\bibnamefont {Noel}}, \ and\
  \bibinfo {author} {\bibfnamefont {R.}~\bibnamefont {Dovesi}},\ }\href
  {\doibase https://doi.org/10.1016/S0022-3697(03)00219-1} {\bibfield
  {journal} {\bibinfo  {journal} {Journal of Physics and Chemistry of Solids}\
  }\textbf {\bibinfo {volume} {64}},\ \bibinfo {pages} {2183} (\bibinfo {year}
  {2003})}\BibitemShut {NoStop}%
\bibitem [{\citenamefont {Kyung}\ \emph {et~al.}(2021)\citenamefont {Kyung},
  \citenamefont {Kim}, \citenamefont {Kim}, \citenamefont {Kim}, \citenamefont
  {Kim}, \citenamefont {Jung}, \citenamefont {Kwon}, \citenamefont {Kim},
  \citenamefont {Bostwick}, \citenamefont {Denlinger}, \citenamefont
  {Yoshida},\ and\ \citenamefont {Kim}}]{Kyung2021}%
  \BibitemOpen
  \bibfield  {author} {\bibinfo {author} {\bibfnamefont {W.}~\bibnamefont
  {Kyung}}, \bibinfo {author} {\bibfnamefont {C.~H.}\ \bibnamefont {Kim}},
  \bibinfo {author} {\bibfnamefont {Y.~K.}\ \bibnamefont {Kim}}, \bibinfo
  {author} {\bibfnamefont {B.}~\bibnamefont {Kim}}, \bibinfo {author}
  {\bibfnamefont {C.}~\bibnamefont {Kim}}, \bibinfo {author} {\bibfnamefont
  {W.}~\bibnamefont {Jung}}, \bibinfo {author} {\bibfnamefont {J.}~\bibnamefont
  {Kwon}}, \bibinfo {author} {\bibfnamefont {M.}~\bibnamefont {Kim}}, \bibinfo
  {author} {\bibfnamefont {A.}~\bibnamefont {Bostwick}}, \bibinfo {author}
  {\bibfnamefont {J.~D.}\ \bibnamefont {Denlinger}}, \bibinfo {author}
  {\bibfnamefont {Y.}~\bibnamefont {Yoshida}}, \ and\ \bibinfo {author}
  {\bibfnamefont {C.}~\bibnamefont {Kim}},\ }\href {\doibase
  10.1038/s41535-020-00306-1} {\bibfield  {journal} {\bibinfo  {journal} {npj
  Quantum Materials}\ }\textbf {\bibinfo {volume} {6}},\ \bibinfo {pages} {5}
  (\bibinfo {year} {2021})}\BibitemShut {NoStop}%
\end{thebibliography}%

\end{document}